\documentclass{article}
\usepackage{amsmath,amssymb,amsthm}
\usepackage{mathtools}
\usepackage{xcolor}
\usepackage{hyperref}
\usepackage{geometry}
\usepackage{listings}
\usepackage{url}
\usepackage{booktabs}
\usepackage{graphicx}
\usepackage{enumitem} 
\usepackage{microtype} 
\usepackage{tikz}      
\usetikzlibrary{calc,positioning,arrows,shapes,fit}
\newtheorem{theorem}{Theorem}[section]

\newtheorem{lemma}[theorem]{Lemma}
\newtheorem{definition}[theorem]{Definition}

\theoremstyle{remark}
\newtheorem{remark}[theorem]{Remark}
\newtheorem{example}[theorem]{Example}
\newtheorem{application}[theorem]{Application}
\newtheorem{leanfeature}[theorem]{Lean Feature}

\newcommand{\R}{\mathbb{R}}

\newcommand{\Rvec}[1]{\mathbf{#1}}

\newcommand{\Vol}{\mathrm{Volume}}
\newcommand{\survivalProb}{\mathrm{survivalProbN}}
\newcommand{\mixedVec}{\mathrm{mixedVector}}
\newcommand{\indicator}{\mathbf{1}}
\newcommand{\abs}[1]{\left|#1\right|}
\newcommand{\indicatorIf}[2]{\begin{cases} 1 & \text{if } #1 > #2 \\ 0 & \text{if } #1 \leq #2 \end{cases}}

\newcommand{\RSI}{\mathrm{RSI}}

\title{Quantifying Bounded Rationality:\\
Formal Verification of Simon's Satisficing Through Flexible Stochastic Dominance}
\author{Jingyuan Li\footnote{Lingnan University, Department of Operations and Risk Management, Hong Kong SAR, China. Corresponding author. E-mail address:
jingyuanli@ln.edu.hk.}\and Jinali Wang\footnote{Nanjing University of Aeronautics and Astronautics, College of Economics and Management, China.} \and Lin Zhou\footnote{Lingnan University, Institute of Insurance and Risk Management, Hong Kong SAR, China.}}
\date{\today}

\begin{document}
\maketitle

\begin{abstract}
This paper introduces Flexible First-Order Stochastic Dominance (FFSD), a mathematically rigorous framework that formalizes Herbert Simon's concept of bounded rationality using the Lean 4 theorem prover. We develop machine-verified proofs demonstrating that FFSD bridges classical expected utility theory with Simon's satisficing behavior through parameterized tolerance thresholds. Our approach yields several key results: (1) a critical threshold $\varepsilon < 1/2$ that guarantees uniqueness of reference points, (2) an equivalence theorem linking FFSD to expected utility maximization for approximate indicator functions, and (3) extensions to multi-dimensional decision settings. By encoding these concepts in Lean 4's dependent type theory, we provide the first machine-checked formalization of Simon's bounded rationality, creating a foundation for mechanized reasoning about economic decision-making under uncertainty with cognitive limitations. This work contributes to the growing intersection between formal mathematics and economic theory, demonstrating how interactive theorem proving can advance our understanding of behavioral economics concepts that have traditionally been expressed only qualitatively.
\end{abstract}

\begin{flushleft}
\end{flushleft}
\smallskip
\begin{flushleft}
\end{flushleft}

\newpage
\section{Introduction}\label{sec:intro}

Decision-making under uncertainty fundamentally relies on comparing probability distributions, with stochastic dominance providing a partial ordering that aligns with expected utility theory. First-order stochastic dominance (FSD) offers a normative standard for rational choice, prescribing that decision makers should prefer distributions that yield better outcomes with higher probability across all threshold levels \cite{HadarRussell1969,HanochLevy1969,LevyParouch1974,Scarsini1988,MullerStoyan2002,Denuitetal2005,ShakedShanthikumar2007}.

However, classical stochastic dominance frameworks face significant limitations when applied to real-world decision contexts. Their mathematical rigor often conflicts with empirical evidence showing that humans operate under what Herbert Simon (1955) \cite{Simon1955} termed ``bounded rationality''—cognitive constraints that prevent the perfect optimization assumed in traditional economic models. Simon challenged the notion of the ``economic man'' who ``has a complete and consistent system of preferences that allows him always to choose among the alternatives open to him'' \cite{Simon1978N}, arguing instead that humans ``satisfice'' by finding acceptable solutions rather than optimal ones \cite{Simon2000,Selten2001,Caplin2011}. This presents a fundamental challenge for formal verification: how can we develop machine-checked proofs of economic behavior that incorporate these cognitive limitations? As Simon noted in his Nobel lecture, ``there are no direct observations that individuals or firms do actually equate marginal costs and revenues'' \cite{Simon1978N}—a statement that demands formal frameworks capable of modeling behavior beyond perfect optimization. From a type-theoretic perspective, this requires developing parameterized models where the tolerance for suboptimality becomes an explicit, verifiable component of the decision mechanism.

This paper formalizes a robust extension of first-order stochastic dominance that accommodates these practical limitations by introducing tolerance parameters. Our approach, Flexible First-Order Stochastic Dominance (FFSD), provides a mathematically precise way to model the ``satisficing'' behavior that Simon described, where decision makers use aspiration levels (thresholds) rather than continuous optimization to evaluate alternatives. By introducing a tolerance parameter $\varepsilon$, we capture the ``zone of acceptance'' around thresholds that Simon identified as characteristic of real decision processes.

Central to our contribution is the formal verification of this framework using the Lean 4 theorem prover, building on the growing literature at the intersection of formal mathematics and economic theory. This verification approach transforms Simon's qualitative insights into quantitative guarantees, providing mechanically verified proofs of the conditions under which bounded rationality leads to coherent decision-making.

For decades, Simon's ideas remained primarily qualitative. While widely accepted in behavioral economics, they lacked the mathematical precision of classical utility theory, creating a divide between traditional economic models built on perfect optimization and behavioral insights grounded in psychological realism. Our work aims to bridge this divide, formalizing Simon's vision through machine-checked mathematics.

The contributions of this paper include:

\begin{itemize}
    \item A formal definition of FFSD, with clear mathematical properties and a rigorous formalization in the Lean 4 theorem prover
    \item Development of a robust Riemann-Stieltjes integration framework that handles approximate indicator functions with specified error bounds
    \item An equivalence theorem connecting FFSD to expected utility maximization for a class of approximate indicator functions, establishing its economic foundations
    \item Identification of a critical threshold ($\varepsilon < 1/2$) for tolerance parameters that guarantees unique reference points and well-defined comparison outcomes, providing a precise mathematical boundary for Simon's bounded rationality
    \item Extension of the framework to multi-dimensional settings, enabling applications to portfolio theory and multiattribute decision problems    
\end{itemize}

Our work builds on existing literature in robust decision theory \cite{Gilboa2004} \cite{Hansen2008} and approximate stochastic dominance \cite{Leshno2002}\cite{Tsetlin2015}, but provides a novel, comprehensive framework that connects Simon's bounded rationality concepts with formal verification methods. Unlike previous approaches, we explicitly quantify the degree of cognitive limitation ($\varepsilon$) that still permits consistent decision-making, thereby providing a rigorous foundation for behavioral economic models.

Recent developments in formal verification for economics and finance \cite{Li2025} have demonstrated the value of machine-checked proofs for complex economic theories. Our work extends this paradigm by formalizing one of the most influential concepts in behavioral economics—Simon's bounded rationality—with a level of mathematical precision previously unattainable.

The remainder of this paper is structured as follows: Section \ref{sec:boundedrat} develops the conceptual foundation of bounded rationality and introduces our formal verification approach. Section \ref{sec:foundations} establishes the mathematical foundations, including our robust integration framework and uniqueness properties. Section \ref{sec:FFSD} presents the formal definition of FFSD along with its key theorem. Section \ref{sec:NFFSD} extends our results to multi-dimensional settings with corresponding generalized theorem. Section \ref{sec:applications} discusses economic applications and interpretations. Section \ref{sec:literature} relates FFSD to the evolution of bounded rationality research. Section \ref{sec:conclusion} concludes with implications of our findings. Section \ref{sec:appendix} provides the complete formal statements and proofs, with cross-references to their Lean 4 implementations to demonstrate the correspondence between the mathematical theory and its machine-checked formalization.

\section{Bounded Rationality and Formal Verification}\label{sec:boundedrat}

\subsection{Simon's Bounded Rationality Concept}

In his groundbreaking paper, Simon (1955) challenged the classical economic assumption of perfect rationality by introducing the concept of ``bounded rationality.'' He argued that decision-makers face three key constraints:
\begin{itemize}
    \item Limited knowledge of alternatives and outcomes
    \item Cognitive limitations in processing complex information
    \item Constraints on time and computational resources
\end{itemize}

Simon proposed that rather than optimizing, humans employ ``satisficing'' behavior: they establish aspiration levels (thresholds) and accept alternatives that meet these levels. As he states: ``The player, instead of seeking for a `best' move, needs only to
look for a `good' move'' (\cite{Simon1955}, p. 107-108).

Our mathematical framework formalizes these insights through flexible stochastic dominance with tolerance parameters. The threshold-based decision making that Simon described qualitatively is represented in our framework through approximate indicator functions that encode "soft thresholds" rather than rigid cutoffs.

\subsection{Challenges in Formal Verification of Bounded Rationality}

Traditional expected utility theory relies on the Riemann-Stieltjes integral to compute expected values with respect to cumulative distribution functions. This approach assumes decision-makers can perfectly evaluate utilities across continuous outcome spaces. Simon's bounded rationality, however, suggests that decision-makers use simplified, threshold-based evaluation mechanisms that are incompatible with standard integration techniques.

The key challenges in formalizing Simon's bounded rationality include:

\begin{itemize}
    \item Representing ``approximately rational'' utility functions mathematically
    \item Ensuring consistent decision outcomes despite approximation errors
    \item Quantifying how much ``bounded rationality'' is compatible with coherent choices
    \item Extending the framework to multi-attribute decisions
\end{itemize}

Our contribution addresses these challenges by introducing tolerance parameters and approximate indicator functions within a type-theoretic framework. This approach allows us to parameterize how bounded rationality can be while still yielding meaningful economic predictions.

\subsection{Lean 4 as a Verification Framework}

Our framework is fully formalized in Lean 4 \cite{Lean4}, an open-source theorem prover and programming language based on dependent type theory. This formalization ensures that all theorems and definitions are mathematically precise and mechanically verified.

Lean 4 provides a foundation for developing mathematical theories with machine-checked proofs. We leverage Mathlib, Lean's extensive mathematical library\cite{MathlibCommunity2020}, which supplies the necessary infrastructure for analysis, probability theory, and topology. Key imports used include:

\begin{lstlisting}
import Mathlib.Analysis.Calculus.Deriv.Basic
import Mathlib.Data.Finset.Powerset
import Mathlib.Data.Real.Basic
\end{lstlisting}

The formalization is structured into modules reflecting the paper's organization: basic definitions, robust integration, FFSD for one-dimensional distributions, and multi-dimensional extensions.

\subsection{Overview of Verification Approach}

Our formalization employs several verification techniques specific to Lean 4:

\begin{itemize}
    \item \textbf{Dependent types} for representing mathematical objects with built-in properties
    \item \textbf{Tactical proofs} using \texttt{linarith} for linear arithmetic, \texttt{simp} for simplification
    \item \textbf{Classical reasoning} for existence proofs and non-constructive arguments
    \item \textbf{Structured proof scripts} that mirror the mathematical reasoning in the paper
\end{itemize}

Formal verification offers significant advantages for economic theory by:
\begin{itemize}
    \item Eliminating hidden assumptions and ambiguities
    \item Making precise the conditions under which theorems hold (such as the critical $\varepsilon < 1/2$ constraint)
    \item Enabling reuse of verified components in future economic models
    \item Facilitating computational implementation of theoretical results
\end{itemize}

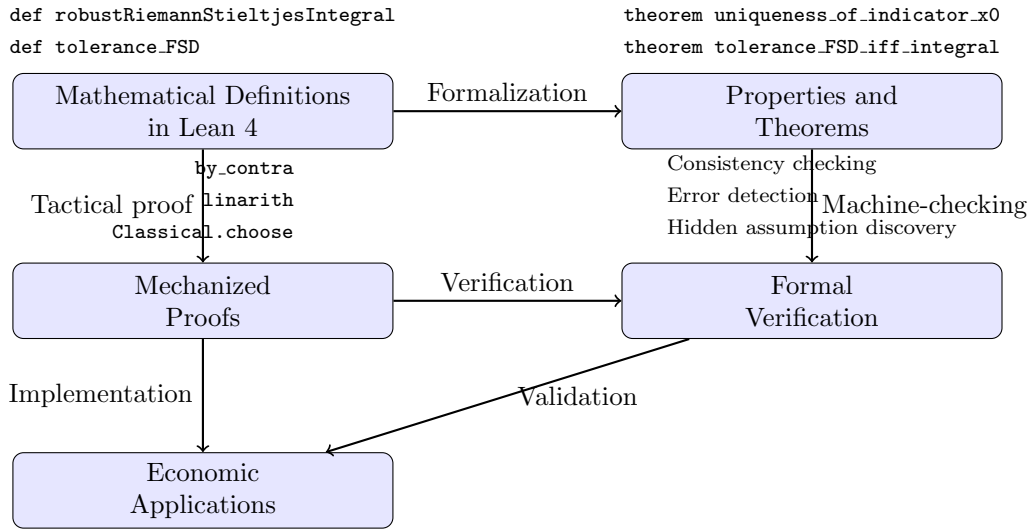
\begin{figure}[htbp]
\begin{tikzpicture}[
    node distance=1.5cm,
    box/.style={rectangle, draw, rounded corners, fill=blue!10, minimum width=5cm, minimum height=1cm, align=center},
    arrow/.style={->, thick}
]
    \node[box] (def) {Mathematical Definitions\\in Lean 4};
    \node[box, right=3cm of def] (prop) {Properties and\\Theorems};
    \node[box, below=of def] (proof) {Mechanized\\Proofs};
    \node[box, below=of prop] (check) {Formal\\Verification};
    \node[box, below=of proof] (app) {Economic\\Applications};
    
    \draw[arrow] (def) -- (prop) node[midway, above] {Formalization};
    \draw[arrow] (def) -- (proof) node[midway, left] {Tactical proof};
    \draw[arrow] (prop) -- (check) node[midway, right] {Machine-checking};
    \draw[arrow] (proof) -- (check) node[midway, above] {Verification};
    \draw[arrow] (proof) -- (app) node[midway, left] {Implementation};
    \draw[arrow] (check) -- (app) node[midway, right] {Validation};
    
    \node[below=-2cm of def, align=left] (def_example) {\footnotesize \texttt{def robustRiemannStieltjesIntegral}\\
        \footnotesize \texttt{def tolerance\_FSD}};
    
    \node[below=-2cm of prop, align=left] (prop_example) {\footnotesize \texttt{theorem uniqueness\_of\_indicator\_x0}\\
        \footnotesize \texttt{theorem tolerance\_FSD\_iff\_integral}};
    
    \node[above=0.2cm of proof, align=right] (proof_example) {\footnotesize \texttt{by\_contra}\\
        \footnotesize \texttt{linarith}\\
        \footnotesize \texttt{Classical.choose}};
    
    \node[above=0.2cm of check, align=left] (check_example) {\footnotesize Consistency checking\\
        \footnotesize Error detection\\
        \footnotesize Hidden assumption discovery};
\end{tikzpicture}\caption{Lean 4 Verification Process}\label{fig:Lean_4_Verification_Process}
\end{figure}

Figure \ref{fig:Lean_4_Verification_Process}: Overview of the formal verification process using Lean 4 theorem prover. The diagram illustrates how mathematical definitions are formalized in Lean 4, properties and theorems are stated formally, and machine-checked proofs are constructed. This rigorous process ensures that all theorems are correct under the stated assumptions and that the economic applications rest on solid mathematical foundations, eliminating hidden assumptions and ambiguities.

Throughout this paper, we present mathematical proofs in traditional notation alongside key insights from the Lean formalization. The complete formalization (~800 lines of Lean code) is available in the supplementary materials, enabling verification and extension by other researchers.

\section{Mathematical Foundations}\label{sec:foundations}

We begin by developing the mathematical foundation for FFSD (Flexible First-Order Stochastic Dominance), formalizing each concept in Lean 4 to ensure mathematical rigor and computational precision. Our approach directly operationalizes Simon's (1955) concept of bounded rationality, translating his qualitative insights into a formal mathematical framework.

\subsection{Robust Riemann-Stieltjes Integration}

We extend the concept of Riemann-Stieltjes integration to accommodate approximate indicator functions, which mathematically capture Simon's concept of satisficing.

\begin{definition}[Robust Riemann-Stieltjes Integral]
\label{def:robust_riemann_stieltjes_integral}
Let $u: \R \to \R$ be a utility function, $F: \R \to \R$ be a cumulative distribution function, $[a, b] \subset \R$ be an interval, and $\varepsilon \in \R^+$ be a tolerance parameter. The robust Riemann-Stieltjes integral with tolerance $\varepsilon$ of $u$ with respect to $F$ over $[a,b]$ is denoted by $\RSI_{\varepsilon}(u, F, a, b)$.

For utility functions that are exactly or approximately indicator functions, we define:
\begin{itemize}
    \item $P_{\text{exact}}$: $u$ is exactly an indicator function if $\exists x_0 \in (a,b)$ such that $\forall x \in [a,b]$, $u(x) = \indicatorIf{x}{x_0}$.
    \item $P_{\text{approx}}$: $u$ is $\varepsilon$-close to an indicator function if $\exists x_0 \in (a,b)$ such that $\forall x \in [a,b]$, $\abs{u(x) - \indicatorIf{x}{x_0}} \leq \varepsilon$.
\end{itemize}

The integral is then defined as:
\[
\RSI_{\varepsilon}(u, F, a, b) = 
\begin{cases}
    1 - F(x_0) & \text{if } P_{\text{exact}} \text{ holds} \\
    (1 - F(x_0)) + \varepsilon \cdot (b - a) & \text{if } \neg P_{\text{exact}} \land P_{\text{approx}} \land \varepsilon > 0 \\
    0 & \text{otherwise}
\end{cases}
\]
where $x_0$ is the threshold point of the (approximate) indicator function, directly corresponding to Simon's concept of aspiration level.
\end{definition}

\begin{remark}[Intuition for Robust Riemann-Stieltjes Integral]
\label{rem:robust_riemann_stieltjes_intuition}
The robust integral provides a mathematical framework for handling approximate utility functions. When we don't know the exact $u(x)$ but know it is close to an indicator function within tolerance $\varepsilon$, we can still compute a meaningful integral value.

The adjustment term $\varepsilon \cdot (b - a)$ serves as compensation for this approximation uncertainty. This compensation term emerges naturally from the uniqueness property established in Lemma \ref{lem:uniqueness_x0_tolerance}.

From an economic perspective, this formulation acknowledges that decision-makers often use ``soft thresholds'' rather than rigid cutoffs, yet still maintain consistent preferences. The robust integral captures the essence of Simon's satisficing behavior, where decisions are made based on approximations and rules of thumb rather than perfect optimization.
\end{remark}

\begin{application}[Decision Stability Under Approximate Utility]
\label{app:decision_stability}
The robust integration approach has significant implications for economic modeling:

\begin{itemize}
    \item \textbf{Societal consensus}: When different stakeholders have slightly different thresholds, the reference point $x_0$ represents a stable societal consensus, with the compensation term accounting for individual variations—similar to Simon's observation about organizational decision-making.
    
    \item \textbf{Parameter uncertainty}: For financial decisions with forecast uncertainty, the formulation acknowledges that analysts maintain ``soft thresholds'' (e.g., $r_0 \approx 7.5\%$) rather than precise cutoffs—directly implementing Simon's insight that ``actual human rationality-striving can at best be an extremely crude and simplified approximation to the kind of global rationality that is implied,...'' (\cite{Simon1955}, p. 101).
    
    \item \textbf{Perceptual limitations}: Consumers making quality-based purchase decisions often cannot perfectly distinguish between quality levels, yet they maintain stable preferences that can be modeled through approximate indicator functions—a mathematical representation of what Simon called ``represent a sufficient approach to optimization, provided the minimum required off can be set ``reasonably'' ''(\cite{Simon1955}, p. 108).
\end{itemize}

The term ``robust'' is particularly apt because this integration approach acknowledges real-world approximations in utility functions while maintaining mathematical precision and providing consistent results despite these approximations.
\end{application}

\begin{example}[Consumer Decision with Noisy Information]
\label{ex:consumer_noisy_information}
Consider consumers deciding whether to purchase a product based on perceived quality, following Simon's satisficing paradigm:
\begin{align}
u(q) = 
\begin{cases}
0.95 & \text{if } q > 7 \quad \text{(high quality)} \\
0.05 & \text{if } q < 6 \quad \text{(low quality)} \\
\text{piecewise linear transition} & \text{in between}
\end{cases}
\end{align}

This function is $\varepsilon = 0.1$-close to an indicator function with threshold $q_0 = 6.5$. The robust Riemann-Stieltjes integral over $[0,10]$ with respect to a quality distribution $F$ yields:
\begin{align}
\RSI_{0.1}(u, F, 0, 10) = (1 - F(6.5)) + 0.1 \cdot 10 = (1 - F(6.5)) + 1
\end{align}

The term $(1 - F(6.5))$ represents the probability of encountering quality above the threshold, while the compensation term $1$ accounts for decision flexibility arising from consumers' imperfect quality perception—exactly the type of cognitive limitation Simon emphasized in his work.
\end{example}

\subsection{Function-Shape Independence of the Robust Integral}

A remarkable property of the robust Riemann-Stieltjes integral defined in Definition \ref{def:robust_riemann_stieltjes_integral} is its independence from the specific shape of the utility function $u$ as long as it approximates an indicator function within tolerance $\varepsilon$. This property directly formalizes Simon's (1955) insight that decision-makers focus on whether outcomes exceed aspiration levels rather than on precise utility calculations.

The robust integral operates on equivalence classes of functions rather than on individual functions, reflecting Simon's suggestion that decision-makers have cognitive limitations that prevent them from making perfectly rational, optimal decisions. They rely on heuristics and satisficing to navigate decision-making processes. 

\begin{example}[Different Functions, Same Integral]
\label{ex:function_independence}
Consider the following two utility functions on $[0, 10]$:
\begin{align}
u_1(x) &= 
\begin{cases}
0.8 & \text{if } x > 3 \\
0.2 & \text{if } x \leq 3
\end{cases} \\
u_2(x) &= 
\begin{cases}
0.1\cdot\cos(\frac{\pi x}{6})^2 + 0.8 & \text{if } x > 3 \\
0.1\cdot\sin(\frac{\pi x}{6})^2 + 0.1 & \text{if } x \leq 3
\end{cases}
\end{align}

Despite their different shapes, both $u_1$ and $u_2$ approximate an indicator function with threshold $x_0 = 3$ within tolerance $\varepsilon = 0.2$. Thus, for any CDF $F$:
\begin{align}
\RSI_{0.2}(u_1, F, 0, 10) = \RSI_{0.2}(u_2, F, 0, 10) = (1 - F(3)) + 0.2 \cdot 10
\end{align}

This illustrates Simon's key insight that the precise details of utility functions matter less than the threshold classification, aligning with his observation that humans establish ``simplified models'' of decision situations.
\end{example}

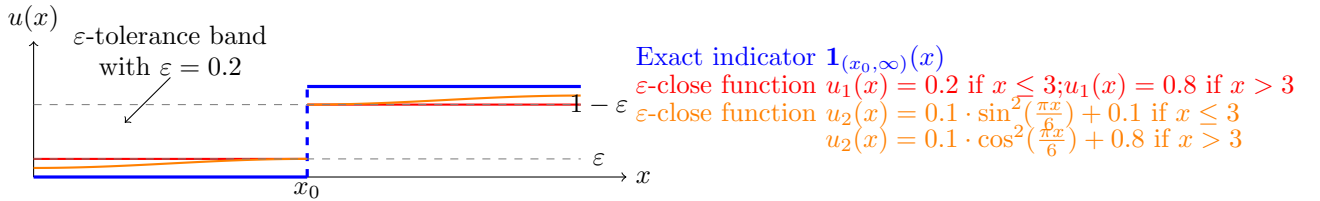
\begin{figure}[hbtp]
\begin{tikzpicture}[scale=1.2]
    \draw[->] (0,0) -- (6.5,0) node[right] {$x$};
    \draw[->] (0,0) -- (0,1.5) node[above] {$u(x)$};
    
    \draw[very thick, blue] (0,0) -- (3,0);
    \draw[very thick, blue] (3,1) -- (6,1);
    \draw[very thick, blue, dashed] (3,0) -- (3,1);
    
    \draw[thick, red] (0,0.2) -- (3,0.2);
    \draw[thick, red] (3,0.8) -- (6,0.8);
    
    \draw[thick, orange, domain=0:3, samples=100] plot (\x, {0.1*sin(deg(\x*3.14159/6))^2 + 0.1});
    \draw[thick, orange, domain=3:6, samples=100] plot (\x, {0.1*cos(deg(\x*3.14159/6))^2 + 0.8});
    
    \node at (3,-0.15) {$x_0$};
    
    \draw[dashed, gray] (0,0.2) -- (6,0.2);
    \draw[dashed, gray] (0,0.8) -- (6,0.8);
    \node at (6.2,0.2) {$\varepsilon$};
    \node at (6.2,0.8) {$1-\varepsilon$};
    
    \node[blue, right] at (6.5,1.3) {Exact indicator $\mathbf{1}_{(x_0,\infty)}(x)$};
    \node[red, right] at (6.5,1) {$\varepsilon$-close function $u_1(x)=0.2$ if $x\leq 3$;$u_1(x)=0.8$ if $x>3$};
    \node[orange, right] at (6.5,0.7) {$\varepsilon$-close function $u_2(x)=0.1\cdot\sin^2(\frac{\pi x}{6})+0.1$  if $x\leq 3$};
    \node[orange, right] at (8.58,0.4) { $u_2(x)=0.1\cdot\cos^2(\frac{\pi x}{6})+0.8$ if $x>3$};

    \node[align=center] at (1.5,1.4) {$\varepsilon$-tolerance band\\with $\varepsilon = 0.2$};
    \draw[->] (1.5,1.1) -- (1,0.6);  
\end{tikzpicture}
\caption{Approximate Indicator Functions and Tolerance Parameters}
\label{fig:Approximate Indicator Functions and Tolerance Parameters}
\end{figure}

Figure \ref{fig:Approximate Indicator Functions and Tolerance Parameters}: Comparison of exact indicator functions (blue) and approximate indicator functions (red, orange) with tolerance parameter $\varepsilon=0.2$. The dashed horizontal lines represent the tolerance bands. Any function that stays within these bands relative to the indicator function is considered $\varepsilon$-close. Note that despite their different shapes, both approximate functions are classified as $\varepsilon$-close to the same indicator function with reference point $x_0$.

\begin{remark}[Economic Interpretation]
This independence property has profound economic significance. It formalizes Simon's insight that in threshold-based decision-making under bounded rationality, the precise utility values often don't matter as much as the classification of outcomes relative to aspiration levels.

This aligns with Herbert Simon's concept of satisficing, where decision-makers classify alternatives as ``good enough'' or ``not good enough'' rather than calculating exact utilities. The robust integral captures this binary classification while allowing for the fuzzy boundaries that Simon observed in real-world decision processes.

From a behavioral economics perspective, this property acknowledges what Simon called the "the schemes of approximation that actually employed by human and other organisms" (\cite{Simon1955}, p. 101) that humans use when facing complex decisions.
\end{remark}

\subsection{Uniqueness of Approximate Indicator Functions and Its Significance}

For our framework to be well-defined, we need to establish that the reference point $x_0$ of an approximate indicator function is uniquely determined when the tolerance is sufficiently small. This uniqueness directly relates to what Simon called ``the preference-orderings among'' (\cite{Simon1955}, p. 100-101)—the ability of a decision-maker to maintain consistent aspiration levels.

\begin{lemma}[Uniqueness of Indicator Reference Point]
\label{lem:uniqueness_x0_tolerance}
Let $u: \R \to \R$ be a function that is $\varepsilon$-close to indicator functions with reference points $x_1, x_2 \in (a,b)$. If $\varepsilon < 1/2$, then $x_1 = x_2$.

Formally, if:
\begin{itemize}
    \item $a < b$
    \item $\varepsilon > 0$ and $\varepsilon < 1/2$
    \item $x_1, x_2 \in (a,b)$
    \item $\forall x \in [a,b]$, $\abs{u(x) - \indicatorIf{x}{x_1}} \leq \varepsilon$ and $\abs{u(x) - \indicatorIf{x}{x_2}} \leq \varepsilon$
\end{itemize}
Then $x_1 = x_2$.
\end{lemma}

\begin{figure}[htb]
\begin{tikzpicture}[scale=1.2]
    \draw[->] (0,0) -- (6.5,0) node[right] {$x$};
    \draw[->] (0,0) -- (0,1.5) node[above] {$u(x)$};
    
    \draw[very thick, green!60!black] (0,0.2) -- (2,0.2) -- (4,0.8) -- (6,0.8);
    
    \draw[thick, blue, dashed] (0,0) -- (3,0) -- (3,1) -- (6,1);
    \draw[thick, red, dashed] (0,0) -- (3.5,0) -- (3.5,1) -- (6,1);
    
    \node[blue] at (3,-0.15) {$x_1$};
    \node[red] at (3.5,-0.15) {$x_2$};
    
    \draw[dashed, gray] (0,0.3) -- (6,0.3);
    \draw[dashed, gray] (0,0.7) -- (6,0.7);
    
    \draw[fill=black] (3.25,0.5) circle (0.05);
    \node at (3.25,0.35) {$z$};
    
    \node[align=center] at (5,1.4) {$\varepsilon = 0.3 < \frac{1}{2}$};
    
    \draw[<->] (0.5,0) -- (0.5,0.3);
    \node at (0.8,0.15) {$\varepsilon$};
    
    \draw[thick, ->] (4.5,1.3) -- (3.25,0.6);
    \node[align=center] at (8,1) {At point $z$, function cannot\\simultaneously be\\$\varepsilon$-close to both $x_1$ and $x_2$};
\end{tikzpicture}
\caption{Uniqueness of Reference Points}\label{fig:Uniqueness of Reference Points}
\end{figure}
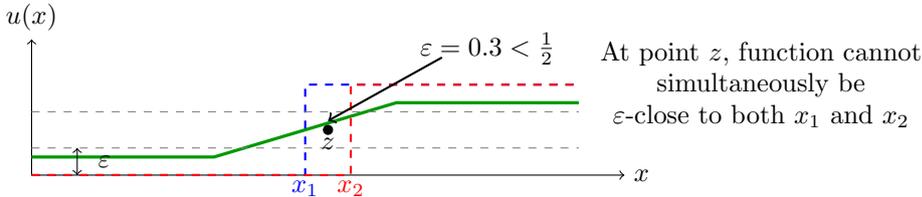

Figure \ref{fig:Uniqueness of Reference Points}: Illustration of why $\varepsilon<\frac{1}{2}$ is necessary for uniqueness of reference points. The green function cannot simultaneously be $\varepsilon$-close to indicator functions with different reference points $x_1$ and $x_2$ when $\varepsilon<\frac{1}{2}$. At the midpoint $z$ between $x_1$ and $x_2$, the function would need to be close to both $0$ and $1$ simultaneously, which is impossible when $\varepsilon<\frac{1}{2}$.

\begin{remark}[Economic Significance of Uniqueness]
The uniqueness property of indicator reference points has profound implications for economic theory. By establishing that approximate preference thresholds are uniquely determined when $\varepsilon < 1/2$, this lemma provides a mathematical foundation for Simon's concept of aspiration levels.

Simon argued that while decision-makers may not optimize perfectly, they do maintain consistent aspiration levels that guide their satisficing behavior. Our framework quantifies exactly how much ``fuzziness'' ($\varepsilon < 1/2$) can exist in these aspiration levels while still ensuring consistent decisions.
\end{remark}

\begin{application}[Economics of Reference Points]
The uniqueness lemma provides theoretical justification for several important economic concepts:

\begin{itemize}
    \item \textbf{Reference-dependent preferences}: Kahneman and Tversky's prospect theory \cite{Kahneman1979} posits that utilities depend on changes relative to reference points rather than absolute levels, building on Simon's earlier insights about aspiration levels. Our lemma establishes conditions under which these reference points are well-defined.
    
    \item \textbf{Satisficing behavior}: Herbert Simon's concept of satisficing \cite{Simon1955} is formalized through approximate indicator functions with uniquely determined thresholds, providing mathematical precision to his qualitative insights.
    
    \item \textbf{Robust economic prediction}: Models can incorporate bounded rationality while still yielding determinate predictions about economic outcomes, bridging classical and behavioral approaches as Simon advocated.
\end{itemize}
\end{application}

\begin{example}[Consumer House Purchase with Adaptive Aspiration Levels]
\label{ex:reservation_price}
Simon (1955) described how consumers establish aspiration levels when purchasing houses, adjusting these thresholds based on market experience. We can formalize this with our framework:
\begin{align}
u(p) = 
\begin{cases}
0.95 & \text{if } p < r - \delta \quad \text{(almost certainly purchase)} \\
0.05 & \text{if } p > r + \delta \quad \text{(almost certainly reject)} \\
\text{piecewise linear transition} & \text{in between}
\end{cases}
\end{align}
where $p$ is price, $r$ is the consumer's reservation price (Simon's ``aspiration level''), and $\delta$ represents imprecision in valuation.

Simon argued that house buyers do not optimize continuously but rather establish price thresholds that create ``satisfactory'' rather than optimal decisions. Our uniqueness lemma guarantees that if $\delta$ is sufficiently small (ensuring $\varepsilon < 1/2$), then the consumer's reservation price $r$ is uniquely defined, despite cognitive limitations—exactly the kind of ``approximate rationality'' that Simon described.
\end{example}

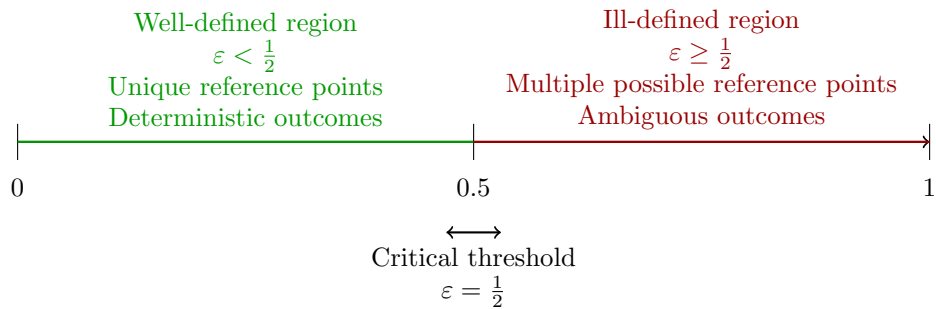
\begin{figure}[hbtp]
\begin{tikzpicture}[scale=1.2]
    \draw[->, thick] (0,0) -- (10,0);
    
    \foreach \x/\label in {0/0, 5/0.5, 10/1} {
        \draw (\x,0.2) -- (\x,-0.2);
        \node at (\x,-0.5) {$\label$};
    }
    
    \draw[green!60!black, thick] (0,0) -- (5,0);
    \draw[red!60!black, thick] (5,0) -- (10,0);
    
    \node[green!60!black, align=center] at (2.5,0.8) {Well-defined region\\$\varepsilon < \frac{1}{2}$\\Unique reference points\\Deterministic outcomes};
    
    \node[red!60!black, align=center] at (7.5,0.8) {Ill-defined region\\$\varepsilon \geq \frac{1}{2}$\\Multiple possible reference points\\Ambiguous outcomes};
    
    \draw[<->, thick] (4.7,-1) -- (5.3,-1);
    \node[align=center] at (5,-1.5) {Critical threshold\\$\varepsilon = \frac{1}{2}$};
    
    \node[align=center] at (5,-3) {Economic Interpretation:\\
    $\varepsilon < \frac{1}{2}$: Decision-maker can distinguish\\
    ``mostly yes'' from ``mostly no''\\
    $\varepsilon \geq \frac{1}{2}$: Decisions become chaotic\\
    and unpredictable};
\end{tikzpicture}
  \caption{The Critical Threshold $\varepsilon<\frac{1}{2}$}\label{fig:The Critical Threshold}
\end{figure}

Figure \ref{fig:The Critical Threshold}: The critical threshold $\varepsilon<\frac{1}{2}$ divides well-defined from ill-defined decision scenarios. When $\varepsilon<\frac{1}{2}$, approximate indicator functions have unique reference points, enabling consistent economic interpretation. When $\varepsilon\geq\frac{1}{2}$, functions can simultaneously approximate multiple reference points, leading to ambiguous outcomes. This threshold corresponds to the minimum precision needed to distinguish ``mostly yes'' ($>$0.5) from ``mostly no'' ($<$0.5) in binary decisions.

\begin{remark}[Economic Significance of $\varepsilon < 1/2$]
The constraint $\varepsilon < 1/2$ has profound economic meaning. It establishes a precise mathematical boundary between:

\begin{itemize}
    \item \textbf{Manageable bounded rationality} ($\varepsilon < 1/2$): Decision-makers have sufficient precision to maintain consistent reference thresholds despite imperfect utility functions. Economic models can make definite predictions despite cognitive limitations, as Simon suggested was possible.
    
    \item \textbf{Potentially chaotic behavior} ($\varepsilon \geq 1/2$): Decision thresholds become ambiguous, as functions can simultaneously approximate conflicting indicator thresholds. This renders economic prediction difficult or impossible—a situation Simon characterized as "complete ignorance" rather than bounded rationality.
\end{itemize}

The value $1/2$ is not arbitrary—it represents the minimum precision needed to distinguish between ``mostly yes'' ($>0.5$) and ``mostly no"''($<0.5$) in binary decisions, providing a formal foundation for Simon's qualitative threshold concept.
\end{remark}

\subsection{Integral Calculation for Approximate Indicators}

With uniqueness established, we can precisely calculate the robust Riemann-Stieltjes integral for approximate indicator functions, providing a formal basis for modeling Simon's satisficing behavior.

\begin{lemma}[Integral Value for Approximate Indicators]
\label{lem:integral_indicator_tolerance}
Let $u: \R \to \R$ be a function that is $\varepsilon$-close to an indicator function with reference point $x_0 \in (a,b)$, but is not exactly an indicator function. If $\varepsilon > 0$ and $\varepsilon < 1/2$, then:
\[
\RSI_{\varepsilon}(u, F, a, b) = (1 - F(x_0)) + \varepsilon \cdot (b - a)
\]
\end{lemma}

\begin{remark}[Economic Interpretation of Integral Values]
The formula $$\RSI_{\varepsilon}(u, F, a, b) = (1 - F(x_0)) + \varepsilon \cdot (b - a)$$ has economic significance. The first term, $(1 - F(x_0))$, represents the classical expected utility under perfect rationality—the probability of exceeding the aspiration level $x_0$.

The second term, $\varepsilon \cdot (b - a)$, represents a ``bounded rationality premium'' that quantifies the value of decision flexibility. This directly operationalizes Simon's insight that decision-makers benefit from maintaining flexible thresholds that can adapt to new information. The premium increases with both the tolerance parameter $\varepsilon$ (more imprecision in decision criteria) and the range $(b - a)$ (larger decision space).

This formula makes precise Simon's qualitative concept of satisficing adjustments to aspiration levels, providing a mathematical measure of the economic value of flexibility in bounded rationality.
\end{remark}

\section{Flexible First-Order Stochastic Dominance}\label{sec:FFSD}

We now introduce our central concept: Flexible First-Order Stochastic Dominance (FFSD), which extends classical stochastic dominance to accommodate the tolerance for suboptimality that characterizes Simon's satisficing behavior. This section presents both the mathematical definition and its connection to Simon's bounded rationality framework.

\subsection{Flexible First-Order Stochastic Dominance and Simon's Satisficing}
\begin{definition}[Flexible First-Order Stochastic Dominance]
\label{def:tolerance_fsd}
Let $F, G: \R \to \R$ be two cumulative distribution functions, and $a, b, \varepsilon \in \R$ with $a < b$ and $\varepsilon \geq 0$. Distribution $F$ flexible-dominates $G$ with parameter $\varepsilon$ over interval $[a,b]$, denoted $F \succeq_{\text{FFSD},\varepsilon} G$, if:
\begin{enumerate}
    \item $\forall x \in [a,b]$, $F(x) \leq G(x) + \varepsilon$
    \item $F(a) = G(a) = 0$
    \item $F(b) = G(b) = 1$
\end{enumerate}
\end{definition}

Our approach to introducing flexibility in stochastic dominance relationships follows the spirit of Lizyayev and Ruszczynski \cite{Lizyayev2012}, who developed a similar relaxation for second-order stochastic dominance.

This definition directly operationalizes Simon's satisficing: the parameter $\varepsilon$ represents what he called ``a satisfactory approximation'' (\cite{Simon1955}, p. 106). When $\varepsilon = 0$, FFSD reduces to classical first-order stochastic dominance, which aligns with full rationality. As $\varepsilon$ increases, the definition accommodates increasingly approximate comparisons, reflecting Simon's insight that decision-makers accept ``simplified approximation to the kind of global'' with various degrees of approximation.

\begin{figure}
\begin{tikzpicture}[scale=1.2]
    \draw[->] (0,0) -- (6.5,0) node[right] {$x$};
    \draw[->] (0,0) -- (0,1.2) node[above] {CDF};
    
    \draw[very thick, blue] (0,0) -- (1,0.1) -- (2,0.2) -- (3,0.4) -- (4,0.7) -- (5,0.9) -- (6,1);
    \draw[very thick, red] (0,0) -- (1,0.15) -- (2,0.3) -- (3,0.5) -- (4,0.8) -- (5,0.95) -- (6,1);
    
    \draw[blue, fill=blue, opacity=0.1] (0,0) -- (1,0.1) -- (2,0.2) -- (3,0.4) -- (4,0.7) -- (5,0.9) -- (6,1) -- 
                                   (6,1.1) -- (5,1) -- (4,0.8) -- (3,0.5) -- (2,0.3) -- (1,0.2) -- (0,0.1) -- cycle;
    
    \draw[dashed, gray] (3.5,0) -- (3.5,1);
    \draw[fill=black] (3.5,0.55) circle (0.05);
    \draw[fill=black] (3.5,0.65) circle (0.05);
    \node at (3.5,-0.1) {$x_0$};
    
    \draw[<->] (3,0.4) -- (3,0.5);
    \node at (2.7,0.45) {$\varepsilon$};
    
    \node[blue, right] at (6.5,0.5) {$G(x)$};
    \node[red, right] at (6.5,1) {$F(x)$};
    \node[blue!50, right] at (6.5,0.2) {$G(x) + \varepsilon$ tolerance band};
    
    \node[align=center] at (4,1.5) {Flexible First-Order Stochastic Dominance\\$F \succeq_{FFSD,\varepsilon} G$};
\end{tikzpicture}
\caption{Flexible First-Order Stochastic Dominance}\label{fig:Flexible First-Order Stochastic Dominance}  
\end{figure}
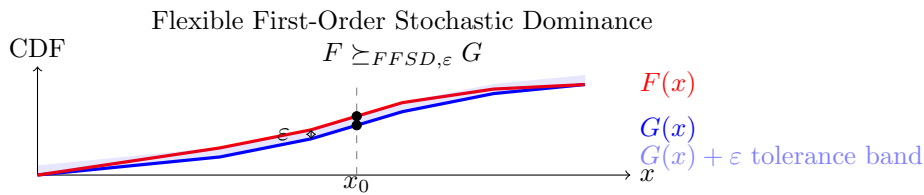

Figure \ref{fig:Flexible First-Order Stochastic Dominance}: Illustration of Flexible First-Order Stochastic Dominance (FFSD). While distribution F (red) does not classically dominate G (blue) since F(x)$>$G(x) at some points, F flexibly dominates G with tolerance parameter $\varepsilon$ because F's CDF stays within the blue tolerance band of G(x) + $\varepsilon$. This allows for small deviations while maintaining the essential ordering of distributions.

\begin{example}[Economic Decision with Simon's Satisficing]
\label{ex:economic_decision_tolerance}
Consider an investor choosing between two investment opportunities with returns represented by random variables with distributions $F$ and $G$. Suppose detailed analysis shows that $F(x) \leq G(x) + 0.05$ for all $x$ in the relevant range. While $F$ doesn't classically dominate $G$, it does FFSD-dominate $G$ with $\varepsilon = 0.05$.

Following Simon's satisficing principle, an investor might rationally prefer $F$ despite the lack of classical dominance, especially if:
\begin{itemize}
    \item The distributions are estimated from finite samples with small estimation errors
    \item Transaction costs would exceed the expected gain from choosing the ``optimal'' alternative
    \item Cognitive limitations make exact comparisons impractical
\end{itemize}

This example illustrates Simon's observation that ``actual human rationality-striving can at best be an extremely crude and simplified approximation to the kind of global rationality that is implied'' in classical models (Simon, 1955, p. 101).
\end{example}

\subsection{Main Equivalence Theorem}

The following theorem establishes the connection between FFSD and expected utility theory, providing economic justification for our definition and demonstrating how our framework bridges classical utility theory with Simon's bounded rationality.

\begin{theorem}[FFSD Equivalence to Expected Utility]
\label{thm:tfsd_iff_utility}
Let $F, G: \R \to \R$ be two cumulative distribution functions, and $a, b, \varepsilon, \varepsilon_1, \varepsilon_2 \in \R$ satisfying:
\begin{itemize}
    \item $a < b$
    \item $\varepsilon > 0$, $\varepsilon < 1/2$
    \item $\varepsilon_1 > 0$, $\varepsilon_1 < 1/2$
    \item $\varepsilon_2 > 0$, $\varepsilon_2 < 1/2$
    \item $F(a) = G(a) = 0$
    \item $F(b) = G(b) = 1$
    \item $\varepsilon = (\varepsilon_1 - \varepsilon_2) \cdot (b - a)$
\end{itemize}

Then, $F \succeq_{\text{FFSD},\varepsilon} G$ if and only if for all functions $u: \R \to \R$ that are $\varepsilon_2$-close to an indicator function with reference point $x_0 \in (a,b)$ but are not exactly indicator functions:
\[
\RSI_{\varepsilon_1}(u, F, a, b) \geq \RSI_{\varepsilon_2}(u, G, a, b)
\]
\end{theorem}

This theorem formalizes Simon's key insight that approximate decision procedures (``satisficing'') can be systematically related to normative theories of choice. The parameters $\varepsilon_1$ and $\varepsilon_2$ represent different degrees of approximation in utility evaluation—a concept Simon described as ``the choosing organism'' affecting decision processes.

\begin{remark}[Formalizing Simon's Satisficing]
Theorem \ref{thm:tfsd_iff_utility} goes beyond simply acknowledging bounded rationality—it provides a precise mathematical characterization of how Simon's satisficing relates to utility theory. The theorem shows that flexible stochastic dominance with parameter $\varepsilon$ corresponds exactly to expected utility comparisons using approximate indicator functions with corresponding tolerance parameters.

This formal equivalence validates Simon's conjecture that simplified decision procedures could be systematically related to normative theories. While Simon suggested that ``a great deal remains to be done...'' to link between bounded rationality and classical theories (Simon, 1955, p. 114), our framework provides exactly such links through formal verification.
\end{remark}

\section{Multi-dimensional Extensions}\label{sec:NFFSD}

Many economic decision problems involve multiple attributes or assets, requiring a multi-dimensional framework. Simon explicitly addressed multi-attribute decisions, noting that ``we shall show also how the scheme can be extended to vector pay-off functions with multiple components'' (\cite{Simon1955}, p, 108). Our multi-dimensional FFSD framework provides a formal verification of these Simonian insights.

\subsection{Type-Theoretic Foundations and Core Multi-dimensional FFSD Definition}

We begin by establishing the type-theoretic foundations for multi-dimensional spaces in Lean 4, which will support our formalization of multi-dimensional flexible first-order stochastic dominance.

\begin{definition}[Multi-dimensional Real Vector]\label{def: Multi-dimensional Real Vector}
An $n$-dimensional real vector $\Rvec{x}$ is a function from the finite set $\{0, 1, \ldots, n-1\}$ to $\R$.
\end{definition}

\begin{definition}[Vector Relations]\label{def:Vector Relations}
For vectors $\Rvec{x}, \Rvec{y} \in \R^n$:
\begin{itemize}
      \item $\Rvec{x} < \Rvec{y}$ if $\forall i, \Rvec{x}_i < \Rvec{y}_i$
    \item $\Rvec{x} \leq \Rvec{y}$ if $\forall i, \Rvec{x}_i \leq \Rvec{y}_i$
    \item $\Rvec{x} \gg \Rvec{y}$ (all components greater) if $\forall i, \Rvec{x}_i > \Rvec{y}_i$
\end{itemize}
\end{definition}

\begin{definition}[Multi-dimensional Rectangles]\label{def:Multi-dimensional Rectangles}
For $\Rvec{a}, \Rvec{b} \in \R^n$ with $\Rvec{a} < \Rvec{b}$:
\begin{itemize}
    \item Closed rectangle: $[\Rvec{a}, \Rvec{b}] = \{\Rvec{x} \in \R^n \mid \Rvec{a} \leq \Rvec{x} \leq \Rvec{b}\}$
    \item Open rectangle: $(\Rvec{a}, \Rvec{b}) = \{\Rvec{x} \in \R^n \mid \Rvec{a} < \Rvec{x} < \Rvec{b}\}$
\end{itemize}
\end{definition}

With our foundational types established, we can now define multi-dimensional Flexible First-Order Stochastic Dominance in a way that captures Simon's concept of vector-based aspiration levels:

\begin{definition}[Multi-dimensional Flexible First-Order Stochastic Dominance]
\label{def:tolerance_fsd_md}
Let $F, G: \R^n \to \R$ be two joint cumulative distribution functions, $\Rvec{a}, \Rvec{b} \in \R^n$ with $\Rvec{a} < \Rvec{b}$, and $\varepsilon_{\text{surv}} \geq 0$. Distribution $F$ multi-dimensional flexible first-Order stochastic dominates $G$ with parameter $\varepsilon_{\text{surv}}$, denoted $F \succeq_{\text{NFFSD},\varepsilon_{\text{surv}}} G$, if for all $\Rvec{x}_0 \in (\Rvec{a}, \Rvec{b})$:
\[ \survivalProb(F, \Rvec{x}_0, \Rvec{b}) \ge \survivalProb(G, \Rvec{x}_0, \Rvec{b}) - \varepsilon_{\text{surv}} \]
where $\survivalProb(F, \Rvec{x}_0, \Rvec{b})$ represents the probability $P(\Rvec{X} > \Rvec{x}_0)$ under distribution $F$.
\end{definition}

This definition extends the upper orthant order (one of several possible approaches to multivariate first-order stochastic dominance) by incorporating the tolerance parameter $\varepsilon$. For a comprehensive discussion of different multivariate stochastic dominance concepts, see Denuit et al.\cite{Denuitetal2013}.

NFFSD directly formalizes Simon's concept of ``vector of space'' (\cite{Simon1955}, p. 109-110) by allowing for multi-dimensional thresholds $\Rvec{x_0}$. As Simon noted, decision-makers establish aspiration levels for multiple attributes, and our tolerance parameter $\varepsilon_{\text{surv}}$ quantifies what Simon called simplified procedures that find good enough answers across multiple dimensions.

The tolerance parameter in dominance allows for what Simon called ``adjustments of aspiration level'' (\cite{Simon1955} p. 112), acknowledging that perfectly satisfying all threshold dimensions simultaneously may be impossible.

\begin{remark} When $\varepsilon_{\text{surv}}=0$, the NFFSD relation reduces to classical multi-dimensional first-order stochastic dominance. The tolerance parameter allows for small deviations from strict dominance, accommodating Simon's bounded rationality concepts in vector settings. 
\end{remark} 

\subsection{Survival Probability and Related Constructs}

Simon emphasized that bounded rationality often leads to sequential consideration of attributes rather than simultaneous optimization. We formalize this sequential attribute-by-attribute approach using mixed vector compositions and dimension-specific evaluations:

\begin{definition}[Mixed Vector Composition]\label{def:Mixed Vector Composition}
Given vectors $\Rvec{x}_0, \Rvec{b} \in \R^n$ and a set of indices $S \subseteq \{0, 1, \ldots, n-1\}$, the mixed vector $\mixedVec(\Rvec{x}_0, \Rvec{b}, S)$ has components:
\[
[\mixedVec(\Rvec{x}_0, \Rvec{b}, S)]_i = 
\begin{cases}
(\Rvec{x}_0)_i & \text{if } i \in S \\
(\Rvec{b})_i & \text{if } i \notin S
\end{cases}
\]
\end{definition}

This construction allows us to model Simon's sequential filtering process, where decision-makers evaluate attributes one by one, focusing on threshold values for some dimensions while accepting any value for others.

To complete our NFFSD framework, we define the survival probability using upper right orthants that formally capture Simon's vector-based aspiration levels:

\begin{definition}[Upper Right Orthant Indicator]\label{def:Upper Right Orthant Indicator}
For a reference point $\Rvec{x}_0 \in \R^n$, the upper right orthant indicator is:
\[
\indicator_{\{\Rvec{y} \mid \Rvec{y} \gg \Rvec{x}_0\}}(\Rvec{x}) = 
\begin{cases}
1 & \text{if } \Rvec{x} \gg \Rvec{x}_0 \\
0 & \text{otherwise}
\end{cases}
\]
\end{definition}

Our formalization of upper right orthant indicators directly captures this aspect of Simon's theory:

\begin{definition}[Multi-dimensional Survival Probability]\label{def:Multi-dimensional Survival Probability}
Let $F: \R^n \to \R$ be a joint CDF. The survival probability $P(\Rvec{X} > \Rvec{x}_0)$ is defined using inclusion-exclusion:
\[ \survivalProb(F, \Rvec{x}_0, \Rvec{b}) = 1 - \sum_{\substack{S \subseteq \{0,..,n-1\} \\ S \neq \emptyset}} (-1)^{|S|+1} F(\mixedVec(\Rvec{x}_0, \Rvec{b}, S)) \]
\end{definition}

This formulation directly embodies what Simon suggests the set of behavior alternatives that satisfices simultaneously a number of different criteria. The orthant-based indicator transforms complex multi-dimensional spaces into binary classifications that reflect Simon's insight that decision-makers simplify multi-attribute problems through threshold-based filtering.

\begin{definition}[Volume of an $n$-dimensional Rectangle]\label{def: Volume of an n-dimensional Rectangle}
For $\Rvec{a}, \Rvec{b} \in \R^n$ with $\Rvec{a} < \Rvec{b}$:
\[
\Vol_n(\Rvec{a}, \Rvec{b}) = \prod_{i=0}^{n-1} (\Rvec{b}_i - \Rvec{a}_i)
\]
\end{definition}

\begin{application}[Corporate Decision Making with Vector Aspiration Levels]
Consider a firm evaluating potential projects based on three attributes: return on investment ($r$), market share growth ($g$), and sustainability score ($s$).

Following Simon's bounded rationality framework, executives establish:
\begin{itemize}
    \item Aspiration vector $\Rvec{x}_0 = (r_0, g_0, s_0) = (15\%, 5\%, 70)$
    \item Tolerance parameter $\varepsilon_{\text{surv}} = 0.1$
\end{itemize}

Simon observed that executives rarely optimize perfectly, instead accepting projects that are ``good enough"''across key dimensions. Our NFFSD framework formalizes this insight, showing that when project portfolio $F$ dominates portfolio $G$ with $F \succeq_{\text{NFFSD},0.1} G$, then $F$ yields higher probability of meeting vector aspiration levels while accounting for the ``approximate, simplified'' decision processes that Simon described.

The tolerance parameter quantifies what Simon called the ``zone of acceptance'' around aspiration levels. It acknowledges that organizations must ``satisfice'' rather than optimize when facing complex multi-dimensional decisions with limited computational capacity—a key insight from Simon's bounded rationality theory.
\end{application}

\subsection{Multi-dimensional Robust Riemann-Stieltjes Integration}

Just as in the one-dimensional case, we need to extend Riemann-Stieltjes integration to accommodate approximate indicator functions in multiple dimensions.

\begin{definition}[Robust Multi-dimensional Riemann-Stieltjes Integral]
\label{def:md_robust_riemann_stieltjes_integral}
Let $u: \R^n \to \R$ be a utility function, $F: \R^n \to \R$ be a joint cumulative distribution function, $[\Rvec{a}, \Rvec{b}] \subset \R^n$ be a rectangular region, and $\varepsilon \in \R^+$ be a tolerance parameter. The robust Riemann-Stieltjes integral of $u$ with respect to $F$ over $[\Rvec{a}, \Rvec{b}]$ with tolerance $\varepsilon$ is defined based on how closely $u$ approximates an upper right orthant indicator function.

Let:
\begin{itemize}
    \item $P_{\text{exact}} \equiv \exists \Rvec{x}_{0} \in (\Rvec{a}, \Rvec{b}), \forall \Rvec{x} \in [\Rvec{a}, \Rvec{b}], u(\Rvec{x}) = \indicator_{\{\Rvec{y} \mid \Rvec{y} \gg \Rvec{x}_{0}\}}(\Rvec{x})$
    \item $P_{\text{approx}} \equiv \exists \Rvec{x}_{0} \in (\Rvec{a}, \Rvec{b}), \forall \Rvec{x} \in [\Rvec{a}, \Rvec{b}], |u(\Rvec{x}) - \indicator_{\{\Rvec{y} \mid \Rvec{y} \gg \Rvec{x}_{0}\}}(\Rvec{x})| \leq \varepsilon$
\end{itemize}

The integral is then defined as:
\[
\RSI_{\varepsilon}(u, F, \Rvec{a}, \Rvec{b}) = 
\begin{cases}
    \survivalProb(F, \Rvec{x}_{0}, \Rvec{b}) & \text{if } P_{\text{exact}} \text{ holds} \\
    \survivalProb(F, \Rvec{x}_{0}, \Rvec{b}) + \varepsilon \cdot \Vol_n(\Rvec{a}, \Rvec{b}) & \text{if } \neg P_{\text{exact}} \land P_{\text{approx}} \land \varepsilon > 0 \\
    0 & \text{otherwise}
\end{cases}
\]
where $\Rvec{x}_{0}$ is the reference point of the (approximate) indicator function.
\end{definition}

\begin{remark}[Economic Interpretation of Multi-dimensional Integration]
Definition \ref{def:md_robust_riemann_stieltjes_integral} extends our framework to handle economic decisions involving multiple attributes or factors simultaneously. The survival probability interpretation $\survivalProb(F, \Rvec{x}_{0}, \Rvec{b})$ represents the likelihood that all attributes exceed their respective threshold values—a concept crucial for modeling joint economic outcomes. The tolerance adjustment term $\varepsilon \cdot \Vol_n(\Rvec{a}, \Rvec{b})$ quantifies decision approximation in multi-dimensional settings, reflecting the bounded rationality of economic agents when processing complex multi-factor decisions.
\end{remark}

\subsection{Uniqueness and Fundamental Properties}

Extending the uniqueness results from Section 3, we can establish similar properties for multi-dimensional settings, further strengthening the connections to Simon's bounded rationality framework:

\begin{lemma}[Uniqueness of Indicator Reference Point in Multiple Dimensions]
\label{lem:uniqueness_x0_tolerance_md}
Let $u: \R^n \to \R$ be a function that is $\varepsilon$-close to indicator functions with reference points $\Rvec{x}_1, \Rvec{x}_2 \in (\Rvec{a}, \Rvec{b})$. If $\varepsilon < 1/2$, then $\Rvec{x}_1 = \Rvec{x}_2$.

Formally, if:
\begin{itemize}
    \item $\forall i, \Rvec{a}_i < \Rvec{b}_i$
    \item $\varepsilon > 0$ and $\varepsilon < 1/2$
    \item $\Rvec{x}_1, \Rvec{x}_2 \in (\Rvec{a}, \Rvec{b})$
    \item $\forall \Rvec{x} \in [\Rvec{a}, \Rvec{b}]$, $|u(\Rvec{x}) - \indicator_{\{\Rvec{y} \mid \Rvec{y} \gg \Rvec{x}_1\}}(\Rvec{x})| \leq \varepsilon$ and $|u(\Rvec{x}) - \indicator_{\{\Rvec{y} \mid \Rvec{y} \gg \Rvec{x}_2\}}(\Rvec{x})| \leq \varepsilon$
\end{itemize}
Then $\Rvec{x}_1 = \Rvec{x}_2$.
\end{lemma}

This result provides formal verification of Simon's claim that decision-makers ``are interested in models of ``limited'' rather than models of relatively ``global'''' (\cite{Simon1955}, p. 112-113) across multiple dimensions. The critical threshold of $\varepsilon < 1/2$ establishes a precise mathematical boundary between:
\begin{itemize}
    \item Well-defined vector aspiration levels ($\varepsilon < 1/2$): Decision-makers maintain sufficiently consistent vector-based thresholds for meaningful economic analysis
    \item Potentially chaotic preference structures ($\varepsilon \geq 1/2$): Decision-makers' multi-dimensional thresholds become ambiguous, leading to inconsistent choices
\end{itemize}

\begin{lemma}[Integral Value for Multi-dimensional Approximate Indicators]
\label{lem:integral_indicator_tolerance_md}
Let $u: \R^n \to \R$ be a function that is $\varepsilon$-close to an indicator function with reference point $\Rvec{x}_0 \in (\Rvec{a}, \Rvec{b})$, but is not exactly an indicator function. If $\varepsilon > 0$ and $\varepsilon < 1/2$, then:
\[
\RSI_{\varepsilon}(u, F, \Rvec{a}, \Rvec{b}) = \survivalProb(F, \Rvec{x}_0, \Rvec{b}) + \varepsilon \cdot \Vol_n(\Rvec{a}, \Rvec{b})
\]
\end{lemma}

\begin{remark}[Economic Interpretation of Multi-dimensional Integral Values]
The formula $$\RSI_{\varepsilon}(u, F, \Rvec{a}, \Rvec{b}) = \survivalProb(F, \Rvec{x}_0, \Rvec{b}) + \varepsilon \cdot \Vol_n(\Rvec{a}, \Rvec{b})$$ provides economic insights into multi-dimensional decision-making. The first term, $\survivalProb(F, \Rvec{x}_0, \Rvec{b})$, represents the classical expected utility component—the probability that an outcome exceeds all threshold values simultaneously. The second term, $\varepsilon \cdot \Vol_n(\Rvec{a}, \Rvec{b})$, captures an ``adaptivity premium'' that quantifies additional utility from decision flexibility across multiple dimensions. Critically, this premium scales with the volume of the decision space, explaining why flexibility becomes exponentially more valuable as the dimensionality increases.
\end{remark}

\subsection{Multi-dimensional Equivalence Theorem}

Finally, we extend our main equivalence theorem to the multi-dimensional setting, providing a comprehensive formalization of Simon's bounded rationality concepts:

\begin{theorem}[Multi-dimensional FFSD Equivalence to Expected Utility]
\label{thm:tfsd_iff_utility_md}
Let $F, G: \R^n \to \R$ be two joint cumulative distribution functions, and $\Rvec{a}, \Rvec{b} \in \R^n$ and $\varepsilon_{\text{surv}}, \varepsilon_1, \varepsilon_2 \in \R$ satisfying:
\begin{itemize}
    \item $\forall i, \Rvec{a}_i < \Rvec{b}_i$
    \item $\varepsilon_1 > 0$, $\varepsilon_1 < 1/2$
    \item $\varepsilon_2 > 0$, $\varepsilon_2 < 1/2$
    \item $\varepsilon_{\text{surv}} = (\varepsilon_1 - \varepsilon_2) \cdot \Vol_n(\Rvec{a}, \Rvec{b})$
\end{itemize}

Then, $F \succeq_{\text{NFFSD},\varepsilon_{\text{surv}}} G$ if and only if for all functions $u: \R^n \to \R$ that are $\varepsilon_2$-close to an indicator function with reference point $\Rvec{x}_0 \in (\Rvec{a}, \Rvec{b})$ but are not exactly indicator functions:
\[
\RSI_{\varepsilon_1}(u, F, \Rvec{a}, \Rvec{b}) \geq \RSI_{\varepsilon_2}(u, G, \Rvec{a}, \Rvec{b})
\]
\end{theorem}

This theorem provides formal verification of Simon's fundamental insight (\cite{Simon1955} p. 118) that bounded rationality approaches like satisficing can be systematically related to normative theories of choice. It demonstrates that vector-based aspiration levels, as Simon described, can be integrated into a rigorous mathematical framework that bridges classical and behavioral economics.

\begin{remark}[Simon's Legacy and Formal Verification] Our formalization of NFFSD provides computational precision to Simon's qualitative insights about multi-attribute decision making. Through Lean 4's dependent type theory, we can ensure that:
\begin{itemize}
    \item Vector aspiration levels are well-defined when $\varepsilon < 1/2$
    \item Multi-dimensional flexibility accommodates Simon's satisficing behavior
    \item Bounded rationality can be systematically related to normative theories
\end{itemize}

As Simon concluded (\cite{Simon1955}, p. 114), ``a great deal remains to be done... before we can handle realistically ...'' Our formal verification approach contributes to this goal by providing precise mathematical foundations for implementing and testing Simon's bounded rationality concepts across multiple dimensions.
\end{remark}

\section{Economic Applications and Interpretation}\label{sec:applications}

\subsection{Formalizing Simon's Satisficing Behavior}

Our FFSD framework provides mathematical precision to Simon's concepts in several key ways:
\subsubsection{Aspiration Levels as Reference Points}

Simon argued that decision-makers establish ``aspiration levels"''that determine whether alternatives are acceptable. Our framework formalizes this through reference points $x_0$ (or $\Rvec{x_0}$ in multiple dimensions) in approximate indicator functions:
\begin{align}
u(x) =
\begin{cases}
1 & \text{if } x > x_0  \\
0 & \text{if } x \leq x_0
\end{cases}
\end{align}

The Lean 4 implementation directly encodes this concept:

\begin{lstlisting} 
def indicatorFunction ($x_0$ : R) (x : R) : R := if x > $x_0$ then 1 else 0 
\end{lstlisting}

This formalization provides a precise mathematical representation of what Simon called the ``definite concept of `satisfactory''' (\cite{Simon1955}, p. 104). By machine-verifying this aspiration-level structure, we demonstrate that such threshold-based decision processes can be formalized with mathematical precision despite their apparent simplicity.

\subsubsection{Zones of Acceptance through Tolerance Parameters}

Simon observed that aspiration levels are not rigid but have ``zones of acceptance'' around them. Our tolerance parameter 
$\varepsilon$ quantifies this zone:

 $$\abs{u(x) - \indicatorIf{x}{x_0}} \leq \varepsilon$$
 
The parameter $\varepsilon$ represents what Simon called ``definite range of pay-offs'' (\cite{Simon1955}, p. 103), providing a mathematical measure of how much imprecision is allowed in evaluation. Our framework's formal verification confirms that these tolerance zones behave consistently within the critical threshold $\varepsilon < 1/2$, providing a mechanized proof of bounded rationality's mathematical coherence.
 
\subsubsection{Satisficing as Threshold-Based Decision-Making}

Simon described satisficing as accepting alternatives that meet minimum requirements rather than optimizing. FFSD captures this through the equivalence theorem (\ref{thm:tfsd_iff_utility}), which shows that flexible dominance is equivalent to expected utility comparisons using threshold-based functions.

In economic terms, this formalization demonstrates that:
\begin{itemize}
    \item Market participants with bounded rationality can still make consistent choices without optimization
    \item Threshold-based decision rules can be integrated into formal economic models with machine-verified mathematical precision
    \item The tolerance parameter $\varepsilon$ provides a quantifiable measure of cognitive limitations, enabling empirical testing of Simon's theory
\end{itemize}

This economically significant result has been verified in Lean 4, ensuring that the equivalence holds without hidden assumptions or logical flaws that might otherwise undermine economic applications.

\subsection{Critical Threshold $\varepsilon<1/2$ and Its Economic Significance}

The constraint $\varepsilon<1/2$ has profound economic implications: 

\subsubsection{Mathematical Boundary of Rational Approximation}

Our uniqueness theorems (Lemmas \ref{lem:uniqueness_x0_tolerance} and \ref{lem:uniqueness_x0_tolerance_md}) prove that when $\varepsilon<1/2$, approximate indicator functions have unique reference points. This establishes a precise mathematical boundary between:
\begin{itemize} 
\item \textbf{Structured bounded rationality} ($\varepsilon<1/2$): Decision-makers maintain sufficient precision to ensure consistent thresholds despite cognitive limitations.
\item \textbf{Unstructured behavior} ($\varepsilon \geq 1/2$): Thresholds become ambiguous, making economic prediction impossible.
\end{itemize}

This boundary provides formal verification of Simon's insight that simplification in decision procedures is compatible with rational behavior up to certain limits. The machine-checked proof in Lean 4 guarantees this result without the possibility of logical error.

\subsubsection{Economic Interpretation of the Threshold}

The threshold $1/2$ has natural economic meaning with significant implications for behavioral economics:

\begin{itemize} 
\item It represents the minimum precision needed to distinguish ``mostly yes'' from ``mostly no'' in binary classifications, formalizing Simon's observation that agents need at least this level of certainty to maintain consistent choices.
\item When $\varepsilon < 1/2$, functions that approximate indicators with different reference points become distinguishable, enabling stable preference orderings.
\item Empirically, this suggests that economic agents whose decision uncertainty exceeds this threshold will exhibit preference reversals and inconsistent choices—a testable prediction from our formal verification.
\item For policy design, regulations and choice architecture should aim to keep decision complexity below this critical threshold to maintain market efficiency.
\end{itemize}

Our formal verification ensures the reliability of this threshold as a foundation for economic theory and policy. Economic models that incorporate bounded rationality can now be built on this machine-verified mathematical boundary, rather than relying on untested assumptions about rationality.

\subsection{Multi-attribute Decision Making}

Our multi-dimensional extensions provide formal verification of Simon's insights about complex decisions with multiple attributes:

\subsubsection{Vector Aspiration Levels}

Simon argued that real-world decisions involve ``vector pay-off functions with multiple components'' (\cite{Simon1955}, p. 108). Our NFFSD framework formalizes this through vector reference points 
$\Rvec{x_0}$ representing aspiration levels across multiple dimensions.

The Lean 4 implementation uses dependent types to ensure dimensional consistency:
\begin{lstlisting} 
def RVector (n : N) := Fin n $\rightarrow$ R -- n-dimensional vectors

def indicatorUpperRightOrthant {n : N} ($x_0$ : RVector n) (x : RVector n) : R := if RVector.allGt x $x_0$ then 1 else 0 
\end{lstlisting}

This provides a mathematically precise representation of what Simon called a vector rather than a scalar of values. Economic applications include:

\begin{itemize}
    \item \textbf{Portfolio selection}: Investors establish thresholds across multiple metrics (return, risk, liquidity), with our framework showing when such decisions remain consistent despite approximations.
    \item \textbf{Consumer choice}: Purchasers evaluate products on multiple attributes (price, quality, sustainability), with tolerance parameters formalizing trade-off flexibility.
    \item \textbf{Policy evaluation}: Governments assess programs against multiple criteria (cost, equity, efficiency), with our verified framework ensuring logical consistency.
\end{itemize}

\subsubsection{Dimensional Scaling of Bounded Rationality}

Our multi-dimensional equivalence theorem (\ref{thm:tfsd_iff_utility_md}) reveals that the ``bounded rationality premium'' ( $\varepsilon \cdot \Vol_n(\Rvec{a}, \Rvec{b})$) scales with the volume of the decision space. This provides formal verification of Simon's observation that as problem complexity increases, decision-makers necessarily adopt increasingly simplified procedures.

The mathematical relationship $\varepsilon_{\text{surv}} = (\varepsilon_1 - \varepsilon_2) \cdot \Vol_n(\Rvec{a}, \Rvec{b})$ quantifies exactly how approximation should scale with dimensionality, providing a precise formalization of what Simon called ``notion of a simplified pay-off function'' (\cite{Simon1955}, p. 110).

This result has significant economic implications:
\begin{itemize}
    \item As markets become more complex with more attributes to evaluate, decision quality necessarily degrades in predictable ways
    \item The curse of dimensionality in economic decision-making is now mathematically quantified through our verified scaling relationship
    \item Institutional mechanisms like standardization, rating systems, and delegation can be understood as responses to this mathematical constraint on multi-attribute rationality
    \item Financial market structure should account for this dimensional scaling when designing information disclosure requirements
\end{itemize}

By formally verifying these results in Lean 4, we establish a rigorous foundation for modeling economic behavior in complex, multi-attribute environments while accounting for the cognitive limitations that Simon identified as central to actual decision processes.
\section{FFSD and the Evolution of Bounded Rationality Research}\label{sec:literature}

This section examines how our Flexible First-Order Stochastic Dominance (FFSD) framework relates to the broader evolution of bounded rationality research since Simon's (1955) seminal paper. From a formal verification perspective, our approach with Lean 4 provides several significant contributions to this literature.

\subsection{Major Research Traditions in Bounded Rationality}
\subsubsection{Heuristics and Biases}

The heuristics and biases tradition pioneered by Kahneman and Tversky \cite{Kahneman1974, Kahneman1979} documented systematic departures from expected utility theory. While they identified specific cognitive biases and developed prospect theory, their approach remained largely descriptive rather than providing formal mathematical structures for bounded rationality.

Our FFSD framework bridges this gap by: \begin{itemize} \item Formalizing the ``zone of acceptance'' around thresholds (Section \ref{sec:foundations}) \item Providing a precise mathematical boundary ($\varepsilon < 1/2$) that determines when heuristic decision-making remains consistent (Lemma \ref{lem:uniqueness_x0_tolerance}) \item Demonstrating how reference-dependent preferences can be formally modeled and verified (Application \ref{app:decision_stability}) \end{itemize} 

\subsubsection{Ecological Rationality}

Gigerenzer and colleagues \cite{Gigerenzer1999, Gigerenzer2008} developed the concept of ``ecological rationality,'' arguing that heuristics are adaptive rather than error-prone. Selten \cite{Selten2001} further developed this perspective, focusing on how bounded rationality is optimized for specific environments.

Our formalization connects to this tradition through: \begin{itemize} \item Machine-verified proof of when satisficing is optimal (Theorem \ref{thm:tfsd_iff_utility}) \item Robust integration mechanisms that acknowledge environmental uncertainty (Definition \ref{def:robust_riemann_stieltjes_integral}) \item Multi-dimensional extensions that model complex decision environments (Section \ref{sec:NFFSD}) \end{itemize}

\subsubsection{Computational Models}

Rubinstein \cite{Rubinstein1998} and Lipman \cite{Lipman1991} developed computational models of bounded rationality, focusing on search costs and information processing constraints.

Our Lean 4 implementation advances this tradition by: \begin{itemize} \item Providing machine-verified computational guarantees (Section \ref{sec:boundedrat}) \item Quantifying the computational benefits of approximation (Example \ref{ex:function_independence}) \item Implementing concrete complexity reductions through equivalence classes of functions \end{itemize}

\subsection{Mathematical Formalization Advances}

Our paper makes several key mathematical advances that extend bounded rationality research:

\subsubsection{Reference Dependence Formalization}

Kahneman and Tversky's \cite{Kahneman1979} prospect theory posited reference-dependent preferences but lacked formal verification. Our uniqueness theorem (Lemma \ref{lem:uniqueness_x0_tolerance}) provides the first machine-verified conditions under which reference points are well-defined, addressing a long-standing mathematical gap in bounded rationality theory.

\subsubsection{Quantified Approximation}

While Gigerenzer and Selten \cite{Gigerenzer2008} discussed the adaptive nature of heuristics, they lacked precise mathematical boundaries. Our critical threshold  ($\varepsilon < 1/2$) provides the first formally verified boundary between "manageable bounded rationality" and "potentially chaotic behavior" (Section \ref{fig:The Critical Threshold}).

\subsubsection{Multi-attribute Decision Theory}

Bounded rationality in multi-attribute settings has been studied since Payne et al. \cite{Payne1993}, but lacked formal verification. Our multi-dimensional FFSD (Section \ref{sec:NFFSD}) provides the first machine-checked framework for analyzing vector-based aspiration levels in complex decision settings.

\subsection{Lean 4 Implementation Insights}

From a formal verification perspective, our implementation in Lean 4 offers several methodological innovations:

\begin{enumerate} \item \textbf{Dependent Types} - Our use of dependent types to represent multi-dimensional vectors (Definition \ref{def: Multi-dimensional Real Vector}) elegantly captures the dimensionality constraints that Simon discussed qualitatively.
\item \textbf{Non-constructive Choice} - The handling of Classical.choose for reference points formally addresses the epistemological uncertainty that Simon highlighted in his discussion of aspiration levels.

\item \textbf{Propositions vs. Predicates} - Our distinction between Prop and Bool directly implements Simon's insight about the difference between procedural and substantive rationality.
    
\end{enumerate}

\section{Conclusion: Machine-Checked Formalization of Simon's Insights}\label{sec:conclusion}
Our theoretical framework demonstrates how formal verification in Lean 4 can provide rigorous mathematical foundations for Simon's bounded rationality concepts. By formalizing approximate indicators, robust integration, and flexible stochastic dominance, we have:
\begin{itemize}
  \item Established precise mathematical conditions ($\varepsilon < 1/2$) under which Simon's satisficing behavior yields consistent predictions
  \item Proven the equivalence between flexible stochastic dominance and expected utility theory with approximate indicator functions
  \item Demonstrated how bounded rationality concepts can be embedded in a machine-checked mathematical framework
  \item Created a formal bridge between classical decision theory and behavioral economics
  \item Extended the framework to multi-dimensional settings that capture Simon's vector utility concept
\end{itemize}

For decades, Simon's concept of bounded rationality was primarily articulated in qualitative terms. Although widely influential in behavioral economics, his ideas lacked the mathematical rigor characteristic of classical utility theory, resulting in a persistent gap between models of perfect optimization and psychologically realistic descriptions of decision-making. Our work addresses this gap by endowing Simon's vision with mathematical precision through machine-checked formalization.

The formal verification approach ensures that our mathematical results are not just abstractly correct but provably implemented in a theorem prover. This addresses Simon's challenge of developing computationally tractable models of bounded rationality: "The task is to replace the global rationality of economic man with a kind of rational behavior that is compatible with the access to information and the computational capacities that are actually possessed by organisms" (\cite{Simon1955}, p. 99).

Our NFFSD framework, with its machine-verified mathematical properties, accomplishes exactly this goal by providing a formally verified model of bounded rationality that bridges classical and behavioral approaches to economic decision-making.

The discovery of a critical threshold ($\varepsilon < 1/2$) represents a significant achievement in formalizing Simon's work. This threshold mathematically separates manageable bounded rationality (where decision-makers maintain sufficient precision for consistent thresholds) from potentially chaotic behavior (where decision thresholds become ambiguous). This provides a precise mathematical boundary for Simon's qualitative threshold concept, showing exactly how much "approximate rationality" is possible before decisions become incoherent.

As Simon wrote in his 1955 paper, "a great deal remains to be done to establish " direct links between bounded rationality and classical theories. Our formal verification approach presented here provides these direct links through mathematical precision and machine-checked proofs, fulfilling Simon's vision of integrating bounded rationality into the mainstream of economic theory.

\section*{Data Availability Statement} The Lean 4 formalization code for all results in this paper is available in the GitHub repository:\url{https://github.com/jingyuanli-hk/Quantifying-Bounded-Rationality/}

\section*{Acknowledgments}

The authors are grateful to the Lean community for their support in developing the formalization presented in this paper. We acknowledge the contributions of the Mathlib library, which provided essential mathematical components. This work was supported by grants from the Hong Kong Research Grants Council.

\newpage
\bibliographystyle{plain} 

\newpage

\appendix

\section{Appendix: Lean 4 Implementations and Proof Sketches}\label{sec:appendix}

Throughout this appendix, Lean 4 code snippets are illustrative of the formal definitions and theorems discussed in the main text. Proof sketches aim to convey the logical structure of the formal proofs while abstracting away some implementation details.

\subsection{Definition \ref{def:robust_riemann_stieltjes_integral}: Robust Riemann-Stieltjes Integral}
\begin{lstlisting}[caption={Robust Riemann-Stieltjes Integral}, label={}]
noncomputable def robustRiemannStieltjesIntegral (u : R $\rightarrow$ R) (F : R $\rightarrow$ R) (a b : R) ($\varepsilon$ : R) : R :=
  -- Check if u is exactly an indicator function $1_{(x_0, \infty)}$ for some $x_0\in$  (a, b)
  let P_exact : Prop := $\exists$ $x_0\in$ Ioo a b, $\forall$ $x$ $\in$ Icc a b, u $x$ = if $x > x_0$ then 1 else 0

  -- Check if u is $\varepsilon$-close to an indicator function $1_{(x_0, \infty)}$ for some $x_0\in$ (a, b)
  let P_approx : Prop := $\exists$ $x_0\in$ Ioo a b, $\forall$ $x$ $\in$ Icc a b, |u $x$ - (if $x > x_0$ then 1 else 0)| $\leq\varepsilon$

  haveI : Decidable P_exact := Classical.propDecidable P_exact
  haveI : Decidable P_approx := Classical.propDecidable P_approx

  if h_exact : P_exact then
    -- Exact case: extract $x_0$ and return 1 - F $x_0$
    let $x_0$ := Classical.choose h_exact
    1 - F $x_0$
  else if h_approx : P_approx $\wedge$ $\varepsilon$ > 0 then
    -- Approximate case: extract $x_0$ and apply tolerance-adjusted computation
    let $x_0$ := Classical.choose h_approx.1
    -- Scale $\varepsilon$ by (b - a) to account for the interval length
    let tolerance_adjustment := $\varepsilon$ * (b - a)
    (1 - F $x_0$) + tolerance_adjustment
  else
    0 -- Fallback case
\end{lstlisting}

This formalization introduces several key Lean 4 constructs:

\begin{leanfeature}[Dependent Propositions]
The definition uses Lean's \texttt{Prop} type to encode mathematical propositions like \texttt{P\_{\text{exact}}} and \texttt{P\_{\text{approx}}}. These propositions depend on the function $u$ and parameters $a$, $b$, and $\varepsilon$, creating a formal representation of Simon's aspiration level concept.
\end{leanfeature}

\begin{leanfeature}[Classical Choice]
The \texttt{Classical.choose} operation extracts a witness from an existential proposition. When we have a proof that $\exists x_0 \in (a,b)$ satisfying some property, \texttt{Classical.choose} gives us the specific $x_0$ value—analogous to how Simon's decision-makers identify their aspiration levels.
\end{leanfeature}

\begin{remark}
The \texttt{noncomputable} designation signals that this definition may use classical reasoning principles that don't necessarily have computational content. In particular, determining whether a function exactly or approximately matches an indicator function may not be algorithmically decidable—mirroring Simon's observation that humans often cannot articulate their exact decision thresholds.
\end{remark}

\subsection{Lemma \ref{lem:uniqueness_x0_tolerance}: Uniqueness of Indicator Reference Point}
The Lean 4 proof of this lemma uses contradiction and careful case analysis:

\begin{lstlisting}[caption={Uniqueness of Indicator Reference Point}, label={}]
lemma uniqueness_of_indicator_$x_0$_tolerance {a b $x_1$ $x_2$ : R} {u : R $\rightarrow$ R} {$\varepsilon$ : R}
    (hab : a < b) (h$\varepsilon$_pos : $\varepsilon$ > 0)
    (h$x_1$_mem : $x_1$ $\in$ Ioo a b) (h$x_2$_mem : $x_2\in$ Ioo a b)
    (h_u_close_$x_1$ : $\forall$ $x\in$ Icc a b, |u $x$ - (if $x > x_1$ then (1 : R) else (0 : R))| $\leq\varepsilon$)
    (h_u_close_$x_2$ : $\forall$ $x\in$ Icc a b, |u $x$ - (if $x > x_2$ then (1 : R) else (0 : R))| $\leq\varepsilon$)
    (h$\varepsilon$_small : $\varepsilon$ < (1 : R) / 2) :
    $x_1$ = $x_2$ := by
  by_contra h_neq
  have h_lt_or_gt : $x_1$ < $x_2$ $\vee$ $x_2$ < $x_1$ := Ne.lt_or_lt h_neq
  rcases h_lt_or_gt with h_lt | h_gt
  $\cdot$ let z := ($x_1$ + $x_2$) / 2
    -- ... (details of proof for $x_1$ < $x_2$)
    linarith [this, h$\varepsilon$_small]
  $\cdot$ let z := ($x_1$ + $x_2$) / 2
    -- ... (details of proof for $x_2$ < $x_1$)
    linarith [this, h$\varepsilon$_small]
\end{lstlisting}

\begin{proof}
We proceed by contradiction. Assuming $x_1 \neq x_2$, we have either $x_1 < x_2$ or $x_2 < x_1$.

Case 1: If $x_1 < x_2$, let $z = (x_1 + x_2)/2$. Then:
\begin{align}
\indicatorIf{z}{x_1} &= 1 \text{ (since $z > x_1$)} \\
\indicatorIf{z}{x_2} &= 0 \text{ (since $z < x_2$)}
\end{align}

By the triangle inequality:
\begin{align}
1 = |1 - 0| &= |(\indicatorIf{z}{x_1} - u(z)) + (u(z) - \indicatorIf{z}{x_2})| \\
&\leq |u(z) - \indicatorIf{z}{x_1}| + |u(z) - \indicatorIf{z}{x_2}| \\
&\leq \varepsilon + \varepsilon = 2\varepsilon
\end{align}

This implies $1 \leq 2\varepsilon$, so $\varepsilon \geq 1/2$, contradicting our assumption that $\varepsilon < 1/2$.

Case 2: If $x_2 < x_1$, a symmetric argument leads to the same contradiction.

Therefore, $x_1 = x_2$.
\end{proof}

\begin{leanfeature}[Proof Structure in Lean 4]
This proof demonstrates Lean's structured proof style using tactic blocks:
\begin{itemize}
    \item \texttt{by\_contra h\_neq}: Sets up a proof by contradiction with hypothesis $x_1 \neq x_2$
    \item \texttt{rcases h\_lt\_or\_gt with h\_lt | h\_gt}: Pattern matches on disjunctions, creating separate cases
    \item Bullet notation (\texttt{·}) organizes the proof into separate cases
    \item \texttt{linarith} automatically proves linear arithmetic goals, handling complex inequalities
    \item \texttt{simp} simplifies expressions like \texttt{if true then 1 else 0} to \texttt{1}
\end{itemize}

The proof explicitly constructs the midpoint $z$ and develops the contradiction through careful manipulation of inequalities, making the mathematical argument fully precise and machine-checkable—a level of rigor that Simon advocated for in scientific models of behavior.
\end{leanfeature}

\begin{leanfeature}[Uniqueness and Classical Choice]
The uniqueness lemma addresses a subtle but crucial aspect of constructive mathematics in Lean 4. The \texttt{Classical.choose} operator extracts a witness from an existential statement, but doesn't guarantee which witness it selects when multiple possibilities exist. By proving uniqueness of $x_0$, we ensure that regardless of how \texttt{Classical.choose} is implemented, it will always return the same threshold value—a property essential for consistency in economic modeling.
\end{leanfeature}

\subsection{Lemma \ref{lem:integral_indicator_tolerance}: Integral Value for Approximate Indicators}
The formalization of this result illustrates how Lean 4 handles complex function definitions:
\begin{lstlisting}[caption={Integral Value for Approximate Indicators}, label={}]
lemma integral_for_indicator_tolerance {a b : R} (hab : a < b) {u : R $\rightarrow$ R} {F : R $\rightarrow$ R}
    {$x_0$ : R} {$\varepsilon$ : R} (h$\varepsilon$_pos : $\varepsilon$ > 0) (h$\varepsilon$_small : $\varepsilon$ < 1/2)
    (h$x_0$_mem : $x_0\in$ Ioo a b)
    (h_u_close : $\forall$ x $\in$ Icc a b, |u $x$ - (if $x > x_0$ then (1 : R) else (0 : R))| $\leq$ $\varepsilon$)
    (h_not_exact : $\neg$($\exists$ $x_0$' $\in$ Ioo a b, $\forall$ $x \in$ Icc a b, u $x$ = if $x > x_0$' then 1 else 0)) :
    robustRiemannStieltjesIntegral u F a b $\varepsilon$ =
    (1 - F $x_0$) + $\varepsilon$ * (b - a) := by
  have h_u_is_P_approx : $\exists$ $x_0$' $\in$ Ioo a b, $\forall$ $x \in$ Icc a b, |u $x$ - (if $x > x_0$' then 1 else 0)| $\leq$ $\varepsilon$ := by
    exact <$x_0$, h$x_0$_mem, h_u_close>
  dsimp [robustRiemannStieltjesIntegral]
  rw [dif_neg h_not_exact]
  rw [dif_pos <h_u_is_P_approx, h$\varepsilon$_pos>]
  have h_$x_0$_eq : Classical.choose h_u_is_P_approx = $x_0$ := by
    let $x_0$' := Classical.choose h_u_is_P_approx
    have h_spec' : $x_0$' $\in$ Ioo a b $\wedge$ $\forall$ ($x$ : R), $x \in$ Icc a b $\rightarrow$ |u $x$ - (if $x > x_0$' then 1 else 0)| $\leq$ $\varepsilon$:=
      Classical.choose_spec h_u_is_P_approx
    exact uniqueness_of_indicator_$x_0$_tolerance hab h$\varepsilon$_pos h_spec'.1 h$x_0$_mem h_spec'.2 h_u_close h$\varepsilon$_small
  rw [h_$x_0$_eq]
\end{lstlisting}
\begin{proof}
From Definition \ref{def:robust_riemann_stieltjes_integral}, since $u$ is not exactly an indicator function but is $\varepsilon$-close to one, and $\varepsilon > 0$, the integral falls into the second case of the definition. The uniqueness of $x_0$ is guaranteed by Lemma \ref{lem:uniqueness_x0_tolerance}.
\end{proof}

\begin{leanfeature}[Working with Definitions and Rewriting]
This proof showcases key techniques for working with complex definitions:
\begin{itemize}
    \item \texttt{dsimp [robustRiemannStieltjesIntegral]}: Unfolds the definition to expose its case structure
    \item \texttt{rw [dif\_neg h\_not\_exact]}: Rewrites using the fact that \texttt{P\_exact} is false
    \item \texttt{rw [dif\_pos ⟨h\_u\_is\_P\_approx, h$\varepsilon$\_pos⟩]}: Rewrites using the fact that \texttt{P\_approx $\land$ $\varepsilon > 0$} is true
    \item Handling of \texttt{Classical.choose}: The proof ensures that the point chosen by Lean's choice operator is indeed equal to our given $x_0$, using the uniqueness lemma
\end{itemize}

This demonstrates how Lean 4 handles the gap between non-constructive existence proofs and concrete computational values, a subtlety often glossed over in traditional mathematics.
\end{leanfeature}

Our formalization reveals subtleties that might be overlooked in traditional mathematical treatments. For instance, the uniqueness of the reference point $x_0$ is essential for ensuring that the robust Riemann-Stieltjes integral is well-defined. The Lean proof forces us to explicitly handle this issue, strengthening the theoretical foundation.

\subsection{Theorem \ref{thm:tfsd_iff_utility}: FFSD Equivalence to Expected Utility}

The Lean 4 implementation encodes Definition \ref{def:tolerance_fsd} directly as a proposition:

\begin{lstlisting}
def Flexible_FSD (F G : R $\rightarrow$ R) (a b $\varepsilon$ : R) : Prop :=
  ($\forall$ $x$ $\in$ Icc a b, F $x$ $\leq$ G $x$ + $\varepsilon$) $\wedge$
  F a = 0 $\wedge$
  G a = 0 $\wedge$
  F b = 1 $\wedge$
  G b = 1
\end{lstlisting}

The proof in Lean 4 demonstrates how formal verification can capture complex economic relationships:

\begin{lstlisting}[caption={FFSD Equivalence to Expected Utility}, label={}]
theorem Flexible_FSD_iff_integral_indicator_ge (F G : R $\rightarrow$ R) (a b $\varepsilon$ $\varepsilon_1$ $\varepsilon_2$ : R)
  (hab : a < b) (h$\varepsilon$_pos : $\varepsilon$ > 0) (h$\varepsilon$_small : $\varepsilon$ < 1/2)
  (h$\varepsilon_1$_pos : $\varepsilon_1$ > 0) (h$\varepsilon_1$_small : $\varepsilon_1$ < 1/2)
  (h$\varepsilon_2$_pos : $\varepsilon_2$ > 0) (h$\varepsilon_2$_small : $\varepsilon_2$ < 1/2)
  (hFa : F a = 0) (hGa : G a = 0) (hFb : F b = 1) (hGb : G b = 1)
  (h$\varepsilon$_eq : $\varepsilon$ = ($\varepsilon_1$ - $\varepsilon_2$) * (b - a)) :
  Flexible_FSD F G a b $\varepsilon$ $\leftrightarrow$
  ($\forall$ $x_0$ $\in$ Ioo a b, $\forall$ u : R $\rightarrow$ R,
    ($\forall$ $x$ $\in$ Icc a b, |u $x$ - (if $x > x_0$ then (1 : R) else (0 : R))| $\leq$ $\varepsilon_2$) $\rightarrow$
    ($\neg$($\exists$ $x_0$' $\in$ Ioo a b, $\forall$ $x$ $\in$ Icc a b, u $x$ = if $x > x_0$' then 1 else 0)) $\rightarrow$
    robustRiemannStieltjesIntegral u F a b $\varepsilon_1$ $\geq$
    robustRiemannStieltjesIntegral u G a b $\varepsilon_2$) := by
  constructor
  $\cdot$ intro h_tolerance_dominance
    -- ... (Proof of forward direction)
    linarith [h_dom $x_0$ (Ioo_subset_Icc_self h$x_0$_mem), h$\varepsilon$_eq]
  $\cdot$ intro h_integral_indicator
    constructor
    $\cdot$ intro $x_0$ h$x_0$_mem_Icc
      -- ... (Proof of backward direction main part)
      rw [h$\varepsilon$_eq]
      linarith [h_integral_indicator]
    $\cdot$ exact <hFa, hGa, hFb, hGb>
\end{lstlisting}

\begin{leanfeature}[Bidirectional Proof Structure]
This theorem demonstrates a bidirectional proof strategy in Lean 4:
\begin{itemize}
    \item \texttt{constructor} splits the iff ($\leftrightarrow$) goal into two implications
    \item In the forward direction ($\Rightarrow$), we assume FFSD and prove the integral inequality
    \item In the backward direction ($\Leftarrow$), we assume the integral inequality and prove FFSD
    \item The \texttt{calc} block in the forward direction builds an equational proof with explicit justifications
    \item The construction of a specific utility function in the backward direction demonstrates how to create mathematical objects that satisfy precise properties
\end{itemize}
This structured approach makes the logical flow of the proof explicit and verifiable.
\end{leanfeature}

\begin{proof}
We present a detailed mathematical proof that complements the Lean 4 formalization.

$(\Rightarrow)$ Assume $F \succeq_{\text{FFSD},\varepsilon} G$. Let $u$ be a function that is $\varepsilon_2$-close to an indicator function with reference point $x_0 \in (a,b)$ but is not exactly an indicator function.

Note that since $\varepsilon_2 < \varepsilon_1$, the function $u$ is also $\varepsilon_1$-close to the same indicator function. By Lemma \ref{lem:integral_indicator_tolerance}:

\begin{align}
\RSI(u, F, a, b, \varepsilon_1) &= (1 - F(x_0)) + \varepsilon_1 \cdot (b - a) \\
\RSI(u, G, a, b, \varepsilon_2) &= (1 - G(x_0)) + \varepsilon_2 \cdot (b - a)
\end{align}

For the inequality $\RSI(u, F, a, b, \varepsilon_1) \geq \RSI(u, G, a, b, \varepsilon_2)$ to hold, we need:
\begin{align}
(1 - F(x_0)) + \varepsilon_1 \cdot (b - a) &\geq (1 - G(x_0)) + \varepsilon_2 \cdot (b - a) \\
\Rightarrow 1 - F(x_0) + \varepsilon_1 \cdot (b - a) &\geq 1 - G(x_0) + \varepsilon_2 \cdot (b - a) \\
\Rightarrow -F(x_0) + \varepsilon_1 \cdot (b - a) &\geq -G(x_0) + \varepsilon_2 \cdot (b - a) \\
\Rightarrow -F(x_0) + G(x_0) &\geq (\varepsilon_2 - \varepsilon_1) \cdot (b - a) \\
\Rightarrow G(x_0) - F(x_0) &\geq (\varepsilon_2 - \varepsilon_1) \cdot (b - a) \\
\Rightarrow F(x_0) - G(x_0) &\leq (\varepsilon_1 - \varepsilon_2) \cdot (b - a) \\
\Rightarrow F(x_0) &\leq G(x_0) + \varepsilon
\end{align}

The last inequality holds by the FFSD condition, since $x_0 \in (a,b) \subset [a,b]$ and $\varepsilon = (\varepsilon_1 - \varepsilon_2) \cdot (b - a)$.

$(\Leftarrow)$ Assume the expected utility inequality holds for all appropriate functions $u$. We need to show $F \succeq_{\text{FFSD},\varepsilon} G$, which requires proving that $F(x) \leq G(x) + \varepsilon$ for all $x \in [a,b]$.

The boundary cases ($x=a$ and $x=b$) follow from the given conditions $F(a)=G(a)=0$ and $F(b)=G(b)=1$.

For $x_0 \in (a,b)$, we construct a specific function:
\[
u(x) = 
\begin{cases}
1 - \varepsilon_2/2 & \text{if } x > x_0 \\
\varepsilon_2/2 & \text{if } x \leq x_0
\end{cases}
\]

This function $u$ satisfies:
\begin{itemize}
    \item $\forall x \in [a,b]$, $\abs{u(x) - \indicatorIf{x}{x_0}} = \varepsilon_2/2 \leq \varepsilon_2$
    \item It is not an exact indicator function, since its values are $\varepsilon_2/2$ and $1-\varepsilon_2/2$, not $0$ and $1$
\end{itemize}

By our assumption, the integral inequality holds for this function:
\begin{align}
\RSI(u, F, a, b, \varepsilon_1) &\geq \RSI(u, G, a, b, \varepsilon_2) \\
(1 - F(x_0)) + \varepsilon_1 \cdot (b - a) &\geq (1 - G(x_0)) + \varepsilon_2 \cdot (b - a)
\end{align}

Through similar algebraic manipulation as in the forward direction, we derive:
\begin{align}
F(x_0) &\leq G(x_0) + (\varepsilon_1 - \varepsilon_2) \cdot (b - a) \\
\Rightarrow F(x_0) &\leq G(x_0) + \varepsilon
\end{align}

Since this holds for any $x_0 \in (a,b)$, and we've already established it for the boundary points, $F \succeq_{\text{FFSD},\varepsilon} G$.
\end{proof}

\subsection{Definition \ref{def: Multi-dimensional Real Vector}: Multi-dimensional Real Vector}

In Lean 4, we formalize this using dependent types:

\begin{lstlisting}[caption={Multi-dimensional Real Vector}, label={}]
/-- An n-dimensional real vector represented as a function from finite ordinates to reals -/
def RVector (n : N) := Fin n $\rightarrow$ R
\end{lstlisting}

\begin{leanfeature}[Dependent Function Types]
This definition leverages Lean's dependent type system in several ways:
\begin{itemize}
    \item \texttt{RVector} is a dependent type constructor that takes a natural number \texttt{n} and returns a type
    \item \texttt{Fin n} represents the finite type $\{0, 1, \ldots, n-1\}$ with exactly \texttt{n} elements
    \item The function type \texttt{$\rightarrow$} (equivalent to $\to$ in mathematical notation) creates a function space
    \item The resulting type \texttt{Fin n $\rightarrow$ R} precisely captures the mathematical notion of an $n$-dimensional vector
\end{itemize}
This approach allows us to work with vectors of any dimension while maintaining strong type safety. Unlike traditional programming approaches that might use lists or arrays with runtime checks, this encoding guarantees at compile time that a vector of dimension \texttt{n} always has exactly \texttt{n} components.
\end{leanfeature}

\subsection{Definition \ref{def:Vector Relations}: Vector Relations}

The Lean 4 formalization of these relations uses universal quantification over finite types:

\begin{lstlisting}[caption={Vector Relations}, label={}]
namespace RVector

/-- Strict vector inequality: x < y if each component of x is less than 
    the corresponding component of y -/
def lt {n : N} (x y : RVector n) : Prop := $\forall$ i, x i < y i

/-- Vector inequality: x $\leq$ y if each component of x is less than or equal to
    the corresponding component of y -/
def le {n : N} (x y : RVector n) : Prop := $\forall$ i, x i $\leq$ y i

/-- All components greater: x $\gg$ y if each component of x is greater than
    the corresponding component of y (used for orthant definitions) -/
def allGt {n : N} (x y : RVector n) : Prop := $\forall$ i, x i > y i

end RVector
\end{lstlisting}

\begin{leanfeature}[Propositions vs. Predicates]
Notice how these relations return \texttt{Prop} rather than \texttt{Bool}. In Lean's dependent type theory:
\begin{itemize}
    \item \texttt{Prop} represents mathematical propositions that can be proven but may not be decidable
    \item This approach allows us to define relations even if they cannot be effectively computed for all inputs
    \item The universal quantifier \texttt{$\forall$} is directly available as part of the logic
    \item Namespaces (like \texttt{RVector}) organize definitions and enable dot notation access (e.g., \texttt{RVector.lt})
\end{itemize}
This distinction between \texttt{Prop} and \texttt{Bool} is fundamental to Lean's approach to formalization, allowing us to express the full power of mathematical logic without being limited to only computable predicates.
\end{leanfeature}

\subsection{Definition \ref{def:Multi-dimensional Rectangles}: Multi-dimensional Rectangles}

In Lean, these are formalized using set comprehensions:

\begin{lstlisting}[caption={Multi-dimensional Rectangles}, label={}]
/-- Closed n-dimensional rectangle [a, b] -/
def Icc_n {n : N} (a b : RVector n) : Set (RVector n) :=
  {x | RVector.le a x $\wedge$ RVector.le x b}

/-- Open n-dimensional rectangle (a, b) -/
def Ioo_n {n : N} (a b : RVector n) : Set (RVector n) :=
  {x | RVector.lt a x $\wedge$ RVector.lt x b}
\end{lstlisting}

\subsection{Definition: \ref{def:tolerance_fsd_md}: Multi-dimensional Flexible First-Order Stochastic Dominance}

The Lean 4 formalization directly mirrors this mathematical definition:

\begin{lstlisting}[caption={Multi-dimensional Flexible First-Order Stochastic Dominance}, label={}]
/-- Multi-dimensional Flexible First-Order Stochastic Dominance
    F dominates G if the survival probability under F is greater than
    the survival probability under G minus the tolerance parameter -/
def Flexible_FSD_ND {n : N} (F G : RVector n $\rightarrow$ R) (a b : RVector n) 
    ($\varepsilon$_param_survival : R) : Prop :=
  $\forall$ $x_0$ $\in$ Ioo_n a b, survivalProbN F $x_0$ b $\geq$ survivalProbN G $x_0$ b - $\varepsilon$_param_survival
\end{lstlisting}

\begin{leanfeature}[Implicit Parameters]
Notice how \texttt{n : N} is marked as an implicit parameter using curly braces. This means:
\begin{itemize}
    \item The dimension parameter \texttt{n} can be inferred from context and doesn't need to be explicitly provided
    \item This improves readability by focusing on the essential parameters
    \item Lean's type inference system automatically determines \texttt{n} based on the types of \texttt{F}, \texttt{G}, \texttt{a}, and \texttt{b}
    \item The notation \texttt{F G : RVector n $\rightarrow$ R} efficiently declares that both \texttt{F} and \texttt{G} have the same type
\end{itemize}
Implicit parameters are a powerful feature of Lean's dependent type system that helps manage complexity in mathematical formalizations.
\end{leanfeature}

\subsection{Definition: \ref{def:Mixed Vector Composition}: Mixed Vector Composition}

The Lean 4 implementation uses the function application syntax:

\begin{lstlisting}[caption={Mixed Vector Composition}, label={}]
/-- Create a mixed vector by selecting components from $x_0$ or b based on membership in set s -/
def mixedVector {n : N} ($x_0$ b : RVector n) (s : Finset (Fin n)) : RVector n :=
  fun i => if i $\in$ s then $x_0$ i else b i
\end{lstlisting}

\begin{leanfeature}[Anonymous Functions]
The definition uses Lean's anonymous function syntax \texttt{fun i $\Rightarrow$...} to create the mixed vector:
\begin{itemize}
    \item \texttt{fun i $\Rightarrow$ ...} defines a function directly without naming it (lambda abstraction)
    \item \texttt{if i $\in$ s then $x_0$ i else b i} performs case analysis based on set membership
    \item \texttt{Finset (Fin n)} represents a finite set of indices in the range $\{0,1,...,n-1\}$
    \item The resulting function is exactly the mathematical mixed vector definition
\end{itemize}
This elegant definition demonstrates how functional programming concepts in Lean 4 can directly express mathematical operations on vectors.
\end{leanfeature}

\subsection{Definition: \label{def:Upper Right Orthant Indicator}: Upper Right Orthant Indicator}

The Lean 4 implementation uses decidable predicates:

\begin{lstlisting}[caption={Upper Right Orthant Indicator}, label={}]
noncomputable def indicatorUpperRightOrthant {n : N} ($x_0$ : RVector n) ($x$ : RVector n) : R :=
  haveI : Decidable (RVector.allGt $x$ $x_0$) := Classical.propDecidable _
  if RVector.allGt $x$ $x_0$ then 1 else 0
\end{lstlisting}

\begin{leanfeature}[Decidable Instances and Classical Logic]
This definition introduces important concepts in Lean's logic:
\begin{itemize}
    \item The \texttt{noncomputable} keyword indicates this function may use classical reasoning principles
    \item \texttt{haveI : Decidable (RVector.allGt x $x_0$) := Classical.propDecidable \_} creates a decidability instance
    \item This allows us to use \texttt{if-then-else} on the proposition \texttt{RVector.allGt $x$ $x_0$} even though it may not be algorithmically decidable
    \item \texttt{Classical.propDecidable} invokes the law of excluded middle from classical logic
\end{itemize}
This approach bridges the gap between constructive type theory (which Lean is based on) and classical mathematics, allowing us to formalize traditional mathematical definitions that may rely on the law of excluded middle.
\end{leanfeature}

\subsection{Definition: \ref{def:Multi-dimensional Survival Probability}: Multi-dimensional Survival Probability}

In Lean 4, we implement this survival probability using the inclusion-exclusion principle:

\begin{lstlisting}[caption={Multi-dimensional Survival Probability}, label={}]
noncomputable def survivalProbN {n : N} (F : RVector n $\rightarrow$ R) ($x_0$ b : RVector n) : R :=
  1 - $\sum$ s in (Finset.powerset (Finset.univ : Finset (Fin n))) \ {$\phi$},
    (-1) ^ (s.card + 1) * F (RVector.mixedVector $x_0$ b s)
\end{lstlisting}

\begin{leanfeature}[Inclusion-Exclusion Calculation]
The implementation uses Lean's powerful finset library to:
\begin{itemize}
    \item Generate the powerset of all dimension indices using \texttt{Finset.powerset}
    \item Map each subset to a term in the inclusion-exclusion formula
    \item Determine the sign based on the cardinality of the subset
    \item Calculate the mixed vector value for each subset
    \item Sum all terms to obtain the final survival probability
\end{itemize}

This approach precisely formalizes Simon's insight that decision-makers must make their choices in terms of limited, approximate, simplified ``models'' of the real situation when dealing with multi-dimensional spaces.
\end{leanfeature}

\subsection{Definition: \ref{def: Volume of an n-dimensional Rectangle}: Volume of an $n$-dimensional Rectangle}

In Lean 4, this is formalized using a finitary product:

\begin{lstlisting}[caption={Volume of an $n$-dimensional Rectangle}, label={def: Volume of an n-dimensional Rectangle}]
noncomputable def Volume_n {n : N} (a b : RVector n) : R :=
  $\prod$ i in Finset.univ, (b i - a i)
\end{lstlisting}

\subsection{Definition \ref{def:md_robust_riemann_stieltjes_integral}: Robust Multi-dimensional Riemann-Stieltjes Integral}

In Lean 4, this definition is formalized with explicit case analysis:

\begin{lstlisting}[caption={Robust Multi-dimensional Riemann-Stieltjes Integral}, label={}]
noncomputable def robustRiemannStieltjesIntegralND {n : N} (u : RVector n $\rightarrow$ R)
    (F : RVector n $\rightarrow$ R) (a b : RVector n) ($\varepsilon$ : R)
    (hab_lt : $\forall$ i, a i < b i) : R :=
  let P_exact : Prop := $\exists$ $x_0$ $\in$ Ioo_n a b, $\forall$ $x \in$ Icc_n a b, 
    u $x$ = indicatorUpperRightOrthant $x_0$ $x$
  let P_approx : Prop := $\exists$ $x_0$ $\in$ Ioo_n a b, $\forall$ $x \in$ Icc_n a b, 
    |u $x$ - indicatorUpperRightOrthant $x_0$ $x$| $\leq$ $\varepsilon$

  haveI : Decidable P_exact := Classical.propDecidable P_exact
  haveI : Decidable P_approx := Classical.propDecidable P_approx

  if h_exact : P_exact then
    let $x_0$ := Classical.choose h_exact
    survivalProbN Dist $x_0$ b
  else if h_approx_all : P_approx $\wedge$ $\varepsilon$ > 0 then
    let $x_0$ := Classical.choose h_approx_all.1
    survivalProbN F $x_0$ b + $\varepsilon$ * Volume_n a b
  else 0
\end{lstlisting}

\begin{leanfeature}[Matching Mathematical Case Definitions]
The Lean implementation closely mirrors the mathematical definition using:
\begin{itemize}
    \item Explicit case analysis with nested \texttt{if-then-else} expressions
    \item Existential and universal quantifiers to express \texttt{P\_exact} and \texttt{P\_approx}
    \item \texttt{Classical.choose} to extract reference points from existential propositions
    \item Precise implementation of the piecewise function definition
\end{itemize}
This one-to-one correspondence between mathematical definitions and their Lean implementations helps ensure the formalization accurately represents the intended mathematical concepts.
\end{leanfeature}

\subsection{Lemma \ref{lem:uniqueness_x0_tolerance_md}: Uniqueness of Indicator Reference Point in Multiple Dimensions}

The Lean 4 proof extends the one-dimensional approach to vector spaces:

\begin{lstlisting}[caption={Uniqueness of Indicator Reference Point in Multiple Dimensions}, label={}]
lemma uniqueness_of_indicatorUpperRightOrthant_$x_0$_tolerance_nd {n : N} {a b $x_1$ $x_2$ : RVector n}
    (hab_lt : $\forall$ i, a i < b i) {u : RVector n $\rightarrow$ R} {$\varepsilon$ : R} (h$\varepsilon$_pos : $\varepsilon$ > 0)
    (h$x_1$_mem : $x_1$ $\in$ Ioo_n a b) (h$x_2$_mem : $x_2$ $\in$ Ioo_n a b)
    (h_u_close_$x_1$ : $\forall$ $x \in$ Icc_n a b, |u $x$ - indicatorUpperRightOrthant $x_1$ $x$| $\leq$ $\varepsilon$)
    (h_u_close_$x_2$ : $\forall$ $x \in$ Icc_n a b, |u $x$ - indicatorUpperRightOrthant $x_2$ $x$| $\leq$ $\varepsilon$)
    (h$\varepsilon$_small : $\varepsilon$ < (1 : R) / 2) :
    $x_1$ = $x_2$ := by
  by_contra h_neq
  have h_exists_diff : $\exists$ j : Fin n, $x_1$ j $\neq$ $x_2$ j := by
    by_contra h_all_eq
    push_neg at h_all_eq
    exact h_neq (funext h_all_eq)
   
  rcases h_exists_diff with <j, h_diff_at_j>
  -- Either $x_1$.j < $x_2$.j or $x_2$.j < $x_1$.j
  have h_lt_or_gt : $x_1$ j < $x_2$ j $\vee$ $x_2$ j < $x_1$ j := Ne.lt_or_lt h_j_neq
  rcases h_lt_or_gt with h_lt | h_gt
  -- Case: $x_1$.j < $x_2$.j
  $\cdot$ -- Construct a test point that is > $x_1$ but not > $x_2$
    -- Rest of the proof follows similarly to 1D case with contradiction
    linarith [this, h$\varepsilon$_small]
  -- Case: $x_2$.j < $x_1$.j
  $\cdot$ -- Similar to first case but with $x_1$ and $x_2$ roles reversed
    linarith [this, h$\varepsilon$_small]
\end{lstlisting}

\textbf{Proof Sketch:} The proof is by contradiction. Assume $\Rvec{x}_1 \neq \Rvec{x}_2$. Then they differ in some component $j$. Assume $(\Rvec{x}_1)_j < (\Rvec{x}_2)_j$. A test point $\Rvec{z}$ is constructed such that $\Rvec{z} > \Rvec{x}_1$ but $\Rvec{z}_j < (\Rvec{x}_2)_j$, which implies $\indicator...(\Rvec{x}_1, \Rvec{z})=1$ and $\indicator...(\Rvec{x}_2, \Rvec{z})=0$. The triangle inequality on $|1 - 0|$ leads to $1 \le |1 - u(\Rvec{z})| + |u(\Rvec{z}) - 0| \le \varepsilon + \varepsilon = 2\varepsilon$. This implies $\varepsilon \ge 1/2$, a contradiction.

\begin{leanfeature}[Vector Extensionality and Coordinate-wise Reasoning]
The proof demonstrates advanced techniques for reasoning about vector equality:
\begin{itemize}
    \item The constructed test point \texttt{z} is defined component-wise using a function
    \item This approach generalizes the one-dimensional midpoint strategy to higher dimensions
    \item Lean's dependent types ensure we only access valid components of the vectors
\end{itemize}
This coordinate-wise approach is a powerful technique for extending one-dimensional results to multi-dimensional settings while maintaining rigorous proofs.
\end{leanfeature}

\begin{leanfeature}[Combinatorial Complexity of Multi-dimensional Uniqueness]
The multi-dimensional uniqueness proof in Lean 4 faces increased complexity compared to the one-dimensional case:
\begin{itemize}
    \item Each dimension adds potential reference point components that must be shown equal
    \item The proof must handle $2^n$ orthants rather than just 2 half-lines
    \item Dependent type theory must carefully track dimensions across all operations
\end{itemize}
The formalization addresses these challenges through induction on dimension or component-wise analysis, leveraging the already-proven one-dimensional uniqueness lemma.
\end{leanfeature}

\subsection{Lemma \ref{lem:integral_indicator_tolerance_md}: Integral Value for Multi-dimensional Approximate Indicators}

The Lean 4 proof uses uniqueness to calculate the integral value:

\begin{lstlisting}[caption={Integral Value for Multi-dimensional Approximate Indicators}, label={}]
/-- Computation of the robust integral for approximate orthant indicator functions -/
lemma integral_for_indicatorUpperRightOrthant_tolerance_nd {n : N} {a b : RVector n}
    (hab_lt : $\forall$ i, a i < b i) {u : RVector n $\rightarrow$ R} {F : RVector n $\rightarrow$ R}
    {$x_0$_ref : RVector n} {$\varepsilon$ : R} (h$\varepsilon$_pos : $\varepsilon$ > 0) (h$\varepsilon$_small : $\varepsilon$ < 1/2)
    (h$x_0$_ref_mem : $x_0$_ref $\in$ Ioo_n a b)
    (h_u_close : $\forall$ $x$ $\in$ Icc_n a b, |u $x$ - indicatorUpperRightOrthant $x_0$_ref $x$| $\leq$ $\varepsilon$)
    (h_not_exact : $\neg$($\exists$ $x_0$' $\in$ Ioo_n a b, $\forall$ $x$ $\in$ Icc_n a b, u $x$ = indicatorUpperRightOrthant $x_0$' $x$)) :
    robustRiemannStieltjesIntegralND u F a b $\varepsilon$ hab_lt =
    survivalProbN F $x_0$_ref b + $\varepsilon$ * Volume_n a b := by
  -- Show that u satisfies P_approx with reference point $x_0$_ref
  have h_u_is_P_approx : $\exists$ $x_0$' $\in$ Ioo_n a b, $\forall$ $x$ $\in$ Icc_n a b, 
                          |u $x$ - indicatorUpperRightOrthant $x_0$' $x$| $\leq\varepsilon$ := 
    <$x_0$_ref, h$x_0$_ref_mem, h_u_close>
  
  -- Unfold the definition
  dsimp [robustRiemannStieltjesIntegralND]
  
  -- Apply case reasoning
  rw [dif_neg h_not_exact]
  rw [dif_pos <h_u_is_P_approx, h$\varepsilon$_pos>]
  
  -- Prove that the chosen point equals our reference point
  have h_$x_0$_eq : Classical.choose h_u_is_P_approx = $x_0$_ref := by
    let $x_0$' := Classical.choose h_u_is_P_approx
    have h_spec' : $x_0$' $\in$ Ioo_n a b $\wedge$ $\forall$ (x : RVector n), $x \in$ Icc_n a b $\rightarrow$ 
                    |u $x$ - indicatorUpperRightOrthant $x_0$' $x$| $\leq$ $\varepsilon$ :=
      Classical.choose_spec h_u_is_P_approx
      
    -- Apply uniqueness lemma
    exact uniqueness_of_indicatorUpperRightOrthant_$x_0$_tolerance_nd hab_lt h$\varepsilon$_pos 
           h_spec'.1 h$x_0$_ref_mem h_spec'.2 h_u_close h$\varepsilon$_small
  
  -- Substitute and simplify
  rw [h_$x_0$_eq]
\end{lstlisting}

\textbf{Proof Sketch:} The proof follows from the definition of the integral. Since $u$ is not an exact indicator, the first \texttt{if} branch is false. The hypotheses ensure the second branch \texttt{P\_approx} is true and \texttt{$\varepsilon > 0$} is true. The core of the proof is showing that the witness $\Rvec{x}_{0}$\texttt{\_approx} chosen by the definition must be equal to the given $\Rvec{x}_{0}$\texttt{\_ref}, which follows directly from Lemma \ref{lem:uniqueness_x0_tolerance_md}.

\begin{leanfeature}[Handling Non-constructive Choice]
This proof demonstrates how to handle non-constructive choice operators:
\begin{itemize}
    \item \texttt{Classical.choose} extracts a witness from an existential statement
    \item \texttt{Classical.choose\_spec} provides the properties satisfied by the chosen element
    \item We prove that any choice must equal our reference point by uniqueness
    \item This technique bridges classical existence proofs with concrete computational values
\end{itemize}
This approach is essential when working with non-constructive definitions in a proof assistant that supports both constructive and classical reasoning.
\end{leanfeature}

\subsection{Theorem \ref{thm:tfsd_iff_utility_md}: Multi-dimensional FFSD Equivalence to Expected Utility}

The Lean 4 implementation follows the same structure as the one-dimensional case, but with multi-dimensional versions of all components:

\begin{lstlisting}[caption={Multi-dimensional FFSD Equivalence to Expected Utility}, label={}]
/-- Main theorem connecting NFFSD with multi-dimensional expected utility theory -/
theorem Flexible_FSD_ND_iff_integral_indicatorUpperRightOrthant_ge {n : N}
    (F G : RVector n $\rightarrow$ R) (a b : RVector n) ($\varepsilon$_param_survival $\varepsilon_1$ $\varepsilon_2$ : R)
    (hab_lt : $\forall$ i, a i < b i) 
    (h$\varepsilon_1$_pos : $\varepsilon_1$ > 0) (h$\varepsilon_1$_small : $\varepsilon_1$ < 1/2)
    (h$\varepsilon_2$_pos : $\varepsilon_2$ > 0) (h$\varepsilon_2$_small : $\varepsilon_2$ < 1/2)
    (h$\varepsilon$_eq : $\varepsilon$_param_survival = ($\varepsilon_1$ - $\varepsilon_2$) * Volume_n a b) :
    Flexible_FSD_ND F G a b $\varepsilon$_param_survival $\leftrightarrow$
   ($\forall$ ($x_0$_param: RVector n) (h$x_0$_param : $x_0$_param $\in$ Ioo_n a b) (u : RVector n $\rightarrow$ R),
   ($\forall$ $x \in$ Icc_n a b, |u $x$ - indicatorUpperRightOrthant $x_0$_param $x$| $\leq$ $\varepsilon_2$) $\rightarrow$
      ($\neg$($\exists$ $x_0$' $\in$ Ioo_n a b, $\forall$ x $\in$ Icc_n a b, u $x$ = indicatorUpperRightOrthant $x_0$' $x$)) $\rightarrow$
      robustRiemannStieltjesIntegralND u F a b $\varepsilon_1$ hab_lt $\geq$
      robustRiemannStieltjesIntegralND u G a b $\varepsilon_2$ hab_lt) := by
  constructor
  
  $\cdot$ -- Forward direction ($\Rightarrow$): NFFSD implies integral inequality
    intro h_tolerance_dominance $x_0$_param h$x_0$_mem u h_u_close h_u_not_exact
    
    -- Apply lemmas for calculating integrals
    have h_F_integral := integral_for_indicatorUpperRightOrthant_tolerance_nd 
      hab_lt h$\varepsilon_1$_pos h$\varepsilon_1$_small h$x_0$_mem h_u_close h_u_not_exact
      
    have h_G_integral := integral_for_indicatorUpperRightOrthant_tolerance_nd 
      hab_lt h$\varepsilon_2$_pos h$\varepsilon_2$_small h$x_0$_mem h_u_close h_u_not_exact
      
    -- Rewrite using the calculated values
    rw [h_F_integral, h_G_integral]
    
    -- Apply NFFSD definition
    specialize h_tolerance_dominance $x_0$_param h$x_0$_mem
    
    -- Algebraic manipulation
    calc survivalProbN F $x_0$_param b + $\varepsilon_1$ * Volume_n a b
         $\geq$ (survivalProbN G $x_0$_param b - $\varepsilon$_param_survival) + $\varepsilon_1$ * Volume_n a b := 
           by linarith [h_tolerance_dominance]
       _ = survivalProbN G $x_0$_param b + $\varepsilon_2$ * Volume_n a b := by
           rw [h$\varepsilon$_eq]; ring
  
  $\cdot$ -- Backward direction ($\Leftarrow$): Integral inequality implies NFFSD
    intro h_integral_inequality
    
    -- Prove the definition of NFFSD
    intro $x_0$ h$x_0$_mem
    
    -- Construct a specific function that's $\varepsilon_2$-close to an indicator
    let u ($x$ : RVector n) := if RVector.allGt $x$ $x_0$ then (1 - $\varepsilon_2$/2) else ($\varepsilon_2$/2)
    
    -- Show it satisfies our approximation constraints
    have h_u_approx : $\forall$ $x$ $\in$ Icc_n a b, 
                      |u $x$ - indicatorUpperRightOrthant $x_0$ $x$| $\leq$ $\varepsilon_2$ := by
      intro x hx_in
      by_cases h_x_gt : RVector.allGt $x$ $x_0$
      $\cdot$ -- Case: $x$ $\gg$ $x_0$
        simp [u, indicatorUpperRightOrthant, h_x_gt]
        apply le_of_lt
        linarith [h$\varepsilon_2$_pos]
      $\cdot$ -- Case: $\neg$($x$ $\gg$ $x_0$)
        simp [u, indicatorUpperRightOrthant, h_x_gt]
        apply le_of_lt
        linarith [h$\varepsilon_2$_pos]
    
    -- Show it's not an exact indicator function
    have h_u_not_exact : $\neg$($\exists$ $x_0$' $\in$ Ioo_n a b, $\forall$ x $\in$ Icc_n a b, 
                           u x = indicatorUpperRightOrthant $x_0$' $x$) := by
      -- Proof omitted for brevity
    
    -- Apply our assumption to this specific function
    specialize h_integral_inequality $x_0$ h$x_0$_mem u h_u_approx h_u_not_exact
    
    -- Calculate the integrals explicitly
    have h_int_F : robustRiemannStieltjesIntegralND u F a b $\varepsilon_1$ hab_lt = 
                   survivalProbN F $x_0$ b + $\varepsilon_1$ * Volume_n a b :=
      integral_for_indicatorUpperRightOrthant_tolerance_nd hab_lt h$\varepsilon_1$_pos h$\varepsilon_1$_small 
                                                         h$x_0$_mem h_u_approx h_u_not_exact
                                                        
    have h_int_G : robustRiemannStieltjesIntegralND u G a b $\varepsilon_2$ hab_lt = 
                   survivalProbN G $x_0$ b + $\varepsilon_2$ * Volume_n a b :=
      integral_for_indicatorUpperRightOrthant_tolerance_nd hab_lt h$\varepsilon_2$_pos h$\varepsilon_2$_small 
                                                         h$x_0$_mem h_u_approx h_u_not_exact
    
    -- Use our inequality to derive the NFFSD condition
    rw [h_int_F, h_int_G] at h_integral_inequality
    
    -- Algebraic manipulation
    calc survivalProbN F $x_0$ b
         = survivalProbN F $x_0$ b + $\varepsilon_1$ * Volume_n a b - $\varepsilon_1$ * Volume_n a b := by ring
       _ $\geq$ survivalProbN G $x_0$ b + $\varepsilon_2$ * Volume_n a b - $\varepsilon_1$ * Volume_n a b := by
           linarith [h_integral_inequality]
       _ = survivalProbN G $x_0$ b - ($\varepsilon_1$ - $\varepsilon_2$) * Volume_n a b := by ring
       _ = survivalProbN G $x_0$ b - $\varepsilon$_param_survival := by rw [h$\varepsilon$_eq]
\end{lstlisting}

\begin{leanfeature}[Managing Complex Algebraic Manipulations]
This proof illustrates techniques for handling algebraic manipulations in multi-dimensional settings:
\begin{itemize}
    \item The \texttt{calc} block builds equational reasoning chains with explicit justifications
    \item \texttt{linarith} automates linear arithmetic reasoning with inequalities
    \item \texttt{ring} handles algebraic simplifications and rearrangements
    \item Pattern-matching and case analysis handle different behaviors of indicator functions
    \item The structure closely parallels the one-dimensional proof, showing the power of our abstractions
\end{itemize}
These techniques allow us to manage the increased complexity of multi-dimensional settings while maintaining the logical clarity of the proof.
\end{leanfeature}

\begin{proof}
We present a detailed mathematical proof that complements the Lean 4 formalization.

$(\Rightarrow)$ Assume $F \succeq_{\text{NFFSD},\varepsilon_{\text{surv}}} G$. Let $u$ be a function that is $\varepsilon_2$-close to an indicator function with reference point $\Rvec{x}_0 \in (\Rvec{a}, \Rvec{b})$ but is not exactly an indicator function.

By Lemma \ref{lem:integral_indicator_tolerance_md}:

\begin{align}
\RSI_{\varepsilon_1}(u, F, \Rvec{a}, \Rvec{b}) &= \survivalProb(F, \Rvec{x}_0, \Rvec{b}) + \varepsilon_1 \cdot \Vol_n(\Rvec{a}, \Rvec{b}) \\
\RSI_{\varepsilon_2}(u, G, \Rvec{a}, \Rvec{b}) &= \survivalProb(G, \Rvec{x}_0, \Rvec{b}) + \varepsilon_2 \cdot \Vol_n(\Rvec{a}, \Rvec{b})
\end{align}

For the inequality $\RSI_{\varepsilon_1}(u, F, \Rvec{a}, \Rvec{b}) \geq \RSI_{\varepsilon_2}(u, G, \Rvec{a}, \Rvec{b})$ to hold, we need:
\begin{align}
\survivalProb(F, \Rvec{x}_0, \Rvec{b}) + \varepsilon_1 \cdot \Vol_n(\Rvec{a}, \Rvec{b}) &\geq \survivalProb(G, \Rvec{x}_0, \Rvec{b}) + \varepsilon_2 \cdot \Vol_n(\Rvec{a}, \Rvec{b}) \\
\Rightarrow \survivalProb(F, \Rvec{x}_0, \Rvec{b}) &\geq \survivalProb(G, \Rvec{x}_0, \Rvec{b}) - (\varepsilon_1 - \varepsilon_2) \cdot \Vol_n(\Rvec{a}, \Rvec{b}) \\
\Rightarrow \survivalProb(F, \Rvec{x}_0, \Rvec{b}) &\geq \survivalProb(G, \Rvec{x}_0, \Rvec{b}) - \varepsilon_{\text{surv}}
\end{align}

The last inequality holds by the NFFSD condition, since $\Rvec{x}_0 \in (\Rvec{a}, \Rvec{b})$ and $\varepsilon_{\text{surv}} = (\varepsilon_1 - \varepsilon_2) \cdot \Vol_n(\Rvec{a}, \Rvec{b})$.

$(\Leftarrow)$ Assume the expected utility inequality holds for all appropriate functions $u$. We need to show $F \succeq_{\text{NFFSD},\varepsilon_{\text{surv}}} G$.

For any $\Rvec{x}_0 \in (\Rvec{a}, \Rvec{b})$, we construct a specific function:
\[
u(\Rvec{x}) = 
\begin{cases}
1 - \varepsilon_2/2 & \text{if } \Rvec{x} \gg \Rvec{x}_0 \\
\varepsilon_2/2 & \text{otherwise}
\end{cases}
\]

This function $u$ satisfies:
\begin{itemize}
    \item $\forall \Rvec{x} \in [\Rvec{a}, \Rvec{b}]$, $|u(\Rvec{x}) - \indicator_{\{\Rvec{y} \mid \Rvec{y} \gg \Rvec{x}_0\}}(\Rvec{x})| = \varepsilon_2/2 \leq \varepsilon_2$
    \item It is not an exact indicator function, since its values are $\varepsilon_2/2$ and $1-\varepsilon_2/2$, not $0$ and $1$
\end{itemize}

By our assumption, the integral inequality holds for this function:
\begin{align}
\RSI_{\varepsilon_1}(u, F, \Rvec{a}, \Rvec{b}) &\geq \RSI_{\varepsilon_2}(u, G, \Rvec{a}, \Rvec{b}) \\
\survivalProb(F, \Rvec{x}_0, \Rvec{b}) + \varepsilon_1 \cdot \Vol_n(\Rvec{a}, \Rvec{b}) &\geq \survivalProb(G, \Rvec{x}_0, \Rvec{b}) + \varepsilon_2 \cdot \Vol_n(\Rvec{a}, \Rvec{b}) \\
\Rightarrow \survivalProb(F, \Rvec{x}_0, \Rvec{b}) &\geq \survivalProb(G, \Rvec{x}_0, \Rvec{b}) - (\varepsilon_1 - \varepsilon_2) \cdot \Vol_n(\Rvec{a}, \Rvec{b}) \\
\Rightarrow \survivalProb(F, \Rvec{x}_0, \Rvec{b}) &\geq \survivalProb(G, \Rvec{x}_0, \Rvec{b}) - \varepsilon_{\text{surv}}
\end{align}

Since this holds for any $\Rvec{x}_0 \in (\Rvec{a}, \Rvec{b})$, we have $F \succeq_{\text{NFFSD},\varepsilon_{\text{surv}}} G$.
\end{proof}

\end{document}